\documentstyle[pre,aps,epsfig,amssymb]{revtex}
\begin{document}
\draft
%\tighten

\title{Theory of asymmetric non-additive binary hard-sphere mixtures}

\author{R. Roth\thanks{present address: Max-Planck Institut f{\"u}r 
Metallforschung, Heisenbergstrasse 1, D-70569 Stuttgart, Germany and ITAP,
University of Stuttgart, Pfaffenwaldring 57, D-70569 Stuttgart, Germany.} and 
R. Evans}
\address{H.H. Wills Physics Laboratory, University of Bristol, Bristol BS8 1TL,
United Kingdom}
\author{A. A. Louis}
\address{Department of Chemistry, Lensfield Rd, Cambridge CB2 1EW, United
Kingdom}
\maketitle
\begin{abstract}
We show that the formal procedure of integrating out the degrees of freedom 
of the small spheres in a binary hard-sphere mixture works equally well for
non-additive as it does for additive mixtures. For highly asymmetric mixtures 
(small size ratios) the resulting effective Hamiltonian of the one-component 
fluid of big spheres, which consists of an infinite number of many-body 
interactions, should be accurately approximated by truncating after the term 
describing the effective pair interaction. Using a density functional treatment
developed originally for additive hard-sphere mixtures we determine the zero,
one, and two-body contribution to the effective Hamiltonian. We demonstrate 
that even small degrees of positive or negative non-additivity have 
significant effect on the shape of the depletion potential. The second virial 
coefficient $B_2$, corresponding to the effective pair interaction between two
big spheres, is found to be a sensitive measure of the effects of 
non-additivity. The variation of $B_2$ with the density of the small spheres
shows significantly different behavior for additive, slightly positive and 
slightly negative non-additive mixtures. We discuss the possible repercussions
of these results for the phase behavior of binary hard-sphere mixtures and
suggest that measurements of $B_2$ might provide a means of determining the 
degree of non-additivity in real colloidal mixtures.
\end{abstract}

\section{Introduction}

Mixtures of hard spheres play a pivotal role in the statistical mechanics of
liquids. Not only do they provide a realistic reference system for describing
the structure and thermodynamics of simple atomic mixtures and mixture of
colloidal particles, they are also of considerable intrinsic interest. In
particular, investigating the properties of asymmetric binary hard-sphere
mixtures became a topic of much activity when it was recognized by Biben and
Hansen \cite{Biben91} that such athermal mixtures might afford important
examples of pure entropy-driven fluid-fluid phase separation. The most studied
model is that of {\em additive} hard spheres, where the cross diameter
$\sigma_{bs}=(\sigma_{bb}+\sigma_{ss})/2$ and $\sigma_{bb}$ refers to the
big-big and $\sigma_{ss}$ to the small-small diameters. There is now compelling
evidence to suggest that additive mixtures, with sufficiently small size ratios
$\sigma_{ss}/\sigma_{bb}$, do undergo fluid-fluid phase separation but this
transition remains metastable w.r.t. the fluid-solid transition 
\cite{Dijkstra98}. The other well studied model is the so-called Asakura-Oosawa
(AO) model \cite{Asakura54,Asakura58} of a colloid-polymer mixture in which the
colloid-colloid interaction is hard-sphere like, with diameter $\sigma_{cc}$,
and the colloid-polymer interaction is also hard with diameter $\sigma_{cp}$,
whereas the polymer-polymer interaction is zero, i.e. $\sigma_{pp}=0$,
corresponding to ideal interpenetrating coils. The cross diameter
$\sigma_{cp}=(\sigma_{cc}+2 R_g)/2$, where $R_g$ is the radius of gyration of
the polymer. Various approximate theories 
\cite{Gast83,Lekkerkerker92,Dijkstra99} and some simulation studies for 
simplified versions of the AO model \cite{Dijkstra99,Meijer94} showed that
when the size ratio $2 R_g/\sigma_{cc}$ is larger than about $0.35$ fluid-fluid
phase separation is stable w.r.t. the fluid-solid transition. The AO model can
be regarded as an extreme {\em non-additive} mixture with 
$\sigma_{cp}>(\sigma_{cc}+\sigma_{pp})/2$.

One can, of course, investigate mixtures with arbitrary non-additivity and 
recent studies of binary hard-sphere mixtures have indicated that a small
degree of positive non-additivity in the cross diameter $\sigma_{bs}$ might
be sufficient to induce a fluid-fluid transition \cite{Biben97,Dijkstra98b}. 
By employing an effective one-component treatment and a variety of liquid 
state perturbation theories, Louis et. al \cite{Louis00} have demonstrated 
that non-additivity should have a profound effect on both the fluid-fluid and 
fluid-solid transition in highly asymmetric hard-sphere mixtures. However, 
their treatment is based on an empirical approximation \cite{note} for the 
effective (depletion) potential between two big spheres rather than any 
systematic derivation of an effective one-component Hamiltonian for the 
non-additive hard-sphere mixture. In this paper we develop such an approach 
following the path that was trodden in the recent studies of the additive 
mixture \cite{Dijkstra98} and in the special case of the AO model
\cite{Dijkstra99}. We show that the same formal technique of integrating out 
the degrees of freedom of the smaller species, used in Ref.~\cite{Dijkstra98},
applies equally well to non-additive mixtures (see Sec.~\ref{sec:formal}).

We were motivated toward such an approach by the following considerations:
a) treating highly asymmetric mixtures by brute force simulation is beset by
ergodicity problems and slow equilibration when the packing fraction of the
small species is substantial, b) for small size ratios three and higher body
potentials in the effective Hamiltonian do not have a significant effect on
the phase behavior of the additive \cite{Dijkstra98} or of the AO
\cite{Dijkstra99,Meijer94} model, i.e. in both models phase transitions are 
determined primarily by the effective pairwise potential between the big
particles and we expect the same to be true for small size ratios in systems 
with intermediate degrees of non-additivity. c) the effective pairwise
potential which arises in the formal development of the theory is the depletion
potential, introduced into colloid science by Asakura and Oosawa, and now
much studied by theory, simulation \cite{Biben96,Dickman97} and experiment.
Indeed a variety of experimental techniques \cite{experiments} have been
developed to measure the depletion potential between a colloidal particle,
immersed in a sea of small colloids or non-adsorbing polymer, and a wall or
another big colloid. Interpreting the results of such experiments requires a 
reliable theory. Recently we have shown \cite{Roth00,Goetzelmann99} that a
density functional approach (DFT) provides an accurate means of calculating the
depletion potential for additive hard spheres. In the present paper we show
(Sec.~\ref{sec:dft}) that the same DFT approach remains valid for the
non-additive case and can therefore be used to investigate a much wider class
of depletion potentials than one might have suspected a priori. We find that
even very small degrees of non-additivity can have a very pronounced effect on
the shape of the depletion potential, which leads to significant changes of the
magnitude, and sometimes the sign, of the second virial coefficient $B_2$
associated with the total effective potential between two big spheres. Such
changes in $B_2$ may, in turn, have repercussions for the phase behavior, and
may be directly accessible by experiments.

We begin by defining what we mean by non-additivity. A non-additive binary 
hard-sphere mixture is characterized by the diameters (distances of closest 
approach of the centers of the spheres) $\sigma_{bb}$, $\sigma_{ss}$, and 
$\sigma_{bs}$ where the subscripts $b$ and $s$ denote big and small. These 
diameters describe the (pairwise) interaction potentials between two spheres
\begin{equation} \label{pair}
\Phi_{ij}(r) = \left\{
\begin{array}{lr}
\infty & r< \sigma_{ij}\\
0 & \mbox{otherwise},
\end{array}
\right.
\end{equation}
with $i,j \in \{s,b\}$. We follow the usual convention and introduce the 
non-additivity parameter 
$\Delta$ via
\begin{equation} \label{defna}
\sigma_{bs} = \frac{1}{2} (\sigma_{bb}+\sigma_{ss})(1+\Delta)
\end{equation}
and allow $\Delta$ to be positive or negative. Additive hard-sphere mixtures 
have $\Delta=0$. 

\begin{figure}
\centering\epsfig{file=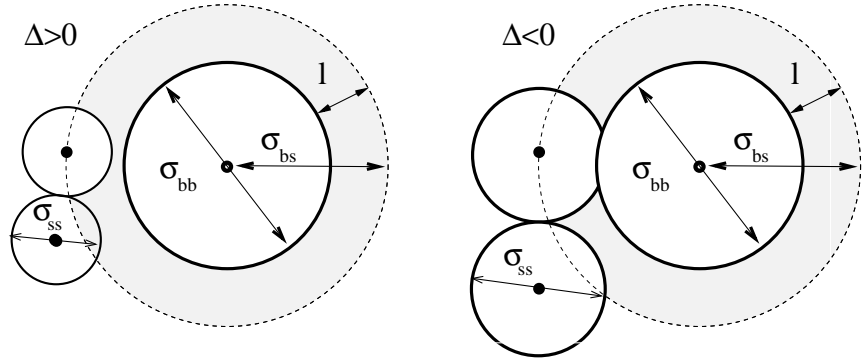,width=12cm,angle=270}
\vspace{0.5cm}
\caption{\label{fig:non-add-A} The centers of the small spheres are excluded
from the shaded depletion layer around each big sphere of diameter 
$\sigma_{bb}$. When changing the non-additivity $\Delta$ by route {\bf (a)}, 
the big-small diameter $\sigma_{bs}$ is kept constant.  For a fixed 
$\sigma_{bb}$ this implies that the depletion layer thickness 
$l= \sigma_{bs}-\sigma_{bb}/2$ is also held constant. The small sphere diameter
$\sigma_{ss}(\Delta)$ is decreased for $\Delta>0$ and increased for $\Delta<0$,
and since the number density $\rho_s^r$ is fixed, the small sphere packing 
fraction $\eta_s^r(\Delta)$ also changes.}
\end{figure}

\begin{figure}
\centering\epsfig{file=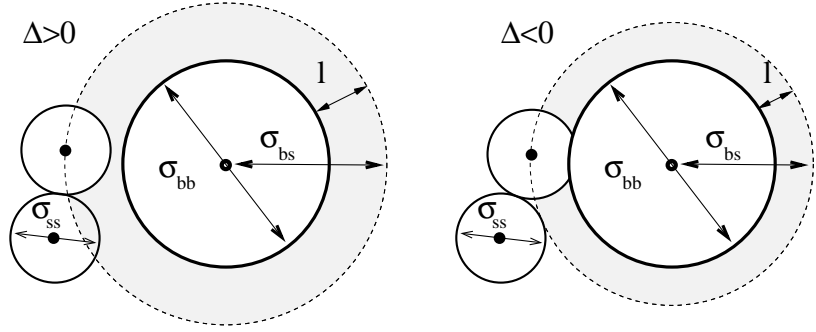,width=12cm,angle=270}
\vspace{0.5cm}
\caption{\label{fig:non-add-B} When changing $\Delta$ by route {\bf (b)}, the 
big-small diameter diameter $\sigma_{bs}$ changes. For a fixed $\sigma_{bb}$ 
this implies that the depletion layer thickness $l=\sigma_{bs}-\sigma_{bb}/2$ 
also changes. The small-particle diameter $\sigma_{ss}$ is kept constant, and 
since $\rho_s^r$ is fixed, the small sphere packing fraction $\eta_s^r$ 
is also constant.}
\end{figure}

In the following we consider two different routes to introducing non-additivity
into a binary hard-sphere mixtures. In both routes we keep the diameter of the
big spheres $\sigma_{bb}$ and the number density of the small spheres in the 
reservoir $\rho_s^r$ fixed but allow for {\bf (a)} changes in the diameter of 
the small spheres $\sigma_{ss}$ while keeping the cross diameter $\sigma_{bs}$ 
constant or {\bf (b)} changes in the cross diameter $\sigma_{bs}$ while keeping
the diameter of the small spheres $\sigma_{ss}$ constant.  These two types of 
changes are illustrated in Fig's~\ref{fig:non-add-A} and \ref{fig:non-add-B} 
respectively.

In the first route {\bf (a)}, which was introduced in an earlier Letter by two
of us \cite{Roth01}, we can smoothly follow a path that connects the additive 
hard-sphere mixture to the Asakura-Oosawa model\cite{Asakura54,Asakura58}.
As mentioned above, the small particles are modeled as ideal gas particles 
with zero diameter $\sigma_{ss}^{AO}\equiv 0$ and $\Delta\equiv\Delta^{AO}>0$, 
keeping the same cross diameter as in the additive case, i.e. we require
\begin{equation}
\sigma_{bs} = \frac{1}{2}(\sigma_{bb}+\sigma_{ss}(\Delta))(1+\Delta) =
\mbox{const.} = \sigma_{bs}^{add} \equiv \frac{1}{2}(\sigma_{bb}+
\sigma_{ss}^{add})
\end{equation}
so that
\begin{equation} \label{sigma}
\sigma_{ss}(\Delta) = \frac{\sigma_{ss}^{add}-\sigma_{bb}~ \Delta}{1+\Delta}.
\end{equation}
The AO limit, defined by $\sigma_{ss}(\Delta^{AO})=0$, implies 
$\Delta^{AO}=\sigma_{ss}^{add}/\sigma_{bb}\equiv q$, the (fixed) size ratio.
Within route {\bf (a)} choosing $\Delta>\Delta^{AO}$ would give rise to 
$\sigma_{ss}(\Delta)<0$ which is unphysical, of course. For a given value of
the non-additivity parameter $\Delta$ and a fixed number density of the
small spheres $\rho_s^r$ it follows that 
their packing fraction $\eta_s^r(\Delta)=\pi \sigma_{ss}^3(\Delta) \rho_s^r/6$ 
also varies with $\Delta$:
\begin{equation} \label{eta}
\eta_s^r(\Delta) = \eta_s^{add} \left(\frac{q - \Delta}{q (1+\Delta)}\right)^3,
\end{equation}
where $\eta_s^{add}\equiv \eta_s^r(0)$. Clearly $\eta_s^r(\Delta)$ decreases 
with increasing $\Delta$ in the range $0<\Delta<\Delta^{AO}=q$. On the other 
hand, for negative non-additivity, $\Delta<0$, $\eta_s^r(\Delta)$ increases 
rapidly with increasing $|\Delta|$ and for studies of the fluid phase we 
should restrict $\eta_s^r(\Delta)<\eta_s^{freez}=0.494$, the value of the 
packing fraction at the bulk freezing transition of hard-spheres.

If we consider route {\bf (b)}, the route used by most previous
authors\cite{Biben97,Dijkstra98b,Louis00}, both $\sigma_{bb}$ and
$\sigma_{ss}$ are kept fixed and the cross diameter $\sigma_{bs}$
becomes a function of $\Delta$ through Eq.~(\ref{defna}). Note that
this case is very different from case {\bf (a)} as the packing
fraction $\eta_s^r$ of the small spheres remains constant through the
whole range of non-additivity and the Asakura-Oosawa limit can only be
reached in the limit of $\rho_s^r\to 0$ in which the depth of the
depletion potential approaches zero as well.

Examples of depletion potential and second virial coefficients calculated from
both routes are given in Sec.~\ref{sec:application}. We conclude in 
Sec.~\ref{sec:discussion} with a summary of our results and a discussion of 
their possible relevance for experiments.

\section{Mapping a binary mixture onto an effective one-component system}

In this section we map the binary mixture of non-additive hard spheres onto an
effective one-component system of big particles. To this end we begin by 
formally integrating out the degrees of freedom of the small spheres and 
determine the form of the effective Hamiltonian of the one-component fluid. 
We demonstrate that each term of the effective Hamiltonian of the non-additive
system can be determined using a theory for an additive mixture. 
In a second step we calculate explicitly the leading contributions to the 
effective Hamiltonian for the binary mixture of non-additive hard spheres 
using a DFT designed for the additive hard-sphere mixture.

\subsection{Formal mapping for additive and non-additive mixtures} 
\label{sec:formal}

We follow the procedure developed in Ref.\cite{Dijkstra98} to formally 
integrate out the degrees of freedom of the small spheres in a homogeneous 
mixture of $N_b$ big and $N_s$ small spheres in a macroscopic volume $V$. The 
Hamiltonian of our mixture is given by
\begin{equation}
H = K + H_{bb} + H_{ss} + H_{bs},
\end{equation}
with $K$ the total kinetic energy of the mixture, leading to a trivial 
contribution to the free energy, and three potential energy contributions: 
\begin{equation}
H_{bb} = \sum_{i<j}^{N_b} \Phi_{bb}({\bf r}_i^b-{\bf r}_j^b),~~~
H_{ss} = \sum_{i<j}^{N_s} \Phi_{ss}({\bf r}_i^s-{\bf r}_j^s),~~~
H_{bs} = \sum_{i=1}^{N_b} \sum_{j=1}^{N_s} \Phi_{bs}({\bf r}_i^b-{\bf r}_j^s)
\end{equation}
where the pairwise hard-sphere interaction potentials $\Phi_{ij}$ are defined 
in Eq.(\ref{pair}). ${\bf r}_i^b$, ${\bf r}_j^s$ denote the coordinates of big 
particle $i$ and small particle $j$, respectively. It was shown in
Ref.~\cite{Dijkstra98} that the thermodynamic potential
$F(N_b,z_s,V)$ of a general binary mixture in the semi-grandcanonical ensemble 
can be written in terms of an effective Hamiltonian $H^{eff}$ via the relation
\begin{equation} \label{defHeff}
\exp(-\beta F) = \mbox{Tr}_b \exp(-\beta H^{eff}),
\end{equation}
where $\beta = 1/k_B T$, and $z_s=\Lambda_s^{-3} \exp(\beta \mu_s)$ is
the fugacity of the (small) species $s$, fixed by the reservoir.
$\mbox{Tr}_\nu$ is shorthand for $1/N_\nu! \Lambda_\nu^{3 N_\nu}$
times the volume integral over the coordinates of species $\nu$.  In
Eq.~(\ref{defHeff})
it denotes the classical trace over the degrees of freedom of the
(big) species $b$.  The effective Hamiltonian $H^{eff}$ is given
\cite{Dijkstra98} by
\begin{equation} \label{Heff}
H^{eff} = H_{bb} + \Omega,
\end{equation}
with $\Omega = \Omega(N_b,z_s,V;\{{\bf r}^b\})$, the grandcanonical potential 
of the fluid of species $s$ subjected to the external potential of a fixed
configuration $\{{\bf r}^b\}$ of $N_b$ particles of species $b$, defined by
\begin{equation}
\exp(-\beta \Omega) = \sum_{N_s=0}^\infty \exp(\beta \mu_s N_s) \mbox{Tr}_s
\exp(-\beta(H_{ss}+H_{bs})). 
\end{equation}
Using a Mayer cluster expansion it was shown \cite{Dijkstra98} that $\Omega$ 
can be written as a sum of terms $\Omega_n$ which describe the simultaneous 
interaction of $n$ particles of species $b$ with the `sea' of species $s$, i.e.
\begin{equation} \label{sum}
\Omega = \sum_{n=0}^{N_b} \Omega_n.
\end{equation}
This result is valid for arbitrary (integrable) pairwise potentials. We 
emphasize that in Eq.~(\ref{Heff}) all direct interactions between species 
$b$, the big particles, are contained in $H_{bb}$ while $\Omega$ describes 
interactions between species $s$, small spheres, and between the big and small
ones. It is precisely this separation into a term which contains just the
big-big interactions and those which contain small-small and big-small
interactions which allows us to calculate the leading terms $\Omega_n$ for 
non-additive hard-sphere mixtures using a theory for an {\em additive} 
mixture.  More specifically, for a given fixed configuration of large 
particles, $\Omega$ is completely determined by the parameters $z_s$, 
$\sigma_{ss}$ and $\sigma_{bs}$. In other words all the terms $\Omega_n$ would 
have an identical form for an additive or a non-additive system. The only 
differences arise in $H_{bb}$, where $\sigma_{bb}$ constrains the possible 
positions of the big particles to $|{\bf r}_i^b - {\bf r}_j^b| > \sigma_{bb}$ 
for all $i,j$.  For the additive case 
$\sigma_{bb} = 2 \sigma_{bs} - \sigma_{ss}$, while in the non-additive case 
$\sigma_{bb}$ can vary more widely depending on the value of $\Delta$.  

\subsubsection{Zero body term $\Omega_0$}

The first term, $\Omega_0$, is the grandcanonical potential of a sea of small 
spheres with fugacity $z_s$ {\em without} any big sphere present and it 
follows that \cite{Dijkstra98}
\begin{equation} \label{Omega0}
\Omega_0(z_s,V) = -p_s(z_s) V,
\end{equation}
with $p_s(z_s)$ the pressure of the reservoir of small spheres. Since this 
term is intrinsic to the small-sphere fluid it is not affected by introducing 
non-additivity.

\subsubsection{One body term $\Omega_1$}

For a homogeneous system the one-body term $\Omega_1$ is of the form
\begin{equation}
\Omega_1(N_b,z_s) = N_b \omega_1(z_s)
\end{equation}
with \cite{Dijkstra98} 
\begin{equation} \label{defw1}
\exp(-\beta \omega_1(z_s))=\langle \exp(-\beta H_{bs}^{(1)})\rangle_{z_s}
\end{equation}
where $H_{bs}^{(n)}$ denotes the interaction between $N_s$ small spheres and
$n \geq 1$ big spheres and the brackets $\langle \dots \rangle_{z_s}$ refer to 
an ensemble average in the reservoir of small spheres. $\omega_1(z_s)$ can be 
identified as the difference in grandcanonical potential between a sea of 
small spheres at fugacity $z_s$ with and without a single big sphere. By
considering the potential distribution theorem and the definition of the
one-body direct correlation function $c_b^{(1)}({\bf r}^b)$ one can show
\cite{Henderson83,Roth00} that this difference in grandcanonical potential can
be expressed as
\begin{equation}
\beta \omega_1(z_s) = - \lim_{\mu_b \to - \infty} c_b^{(1)}(\infty),
\end{equation} 
where $c_b^{(1)}(\infty)$ denotes the direct correlation function of a big 
sphere evaluated in the bulk mixture. The limit $\mu_b \to -\infty$ implies 
that the chemical potential $\mu_b$ of species $b$ is made sufficiently 
negative that only one big sphere is present. $c_b^{(1)}(\infty)$ is 
proportional to the excess (over ideal) chemical potential of species $b$, i.e.
$-c_b^{(1)}(\infty) = \beta \mu_b^{ex}$ and, in general, depends on the density
of both species. However, in the limit $\mu_b \to -\infty$, $c_b^{(1)}(\infty)$
depends only on the fugacity $z_s$. Since only one big sphere is involved
[this is explicit in Eq.~(\ref{defw1})] non-additivity plays no role in 
determining $\Omega_1$. One merely specifies $\sigma_{ss}$ and then 
$\sigma_{bs}$ describes the interaction between small spheres and a fixed big
one. 

As noticed in Ref.~\cite{Dijkstra00}, the one-body term $\omega_1(z_s)$
determines the Henry's law constant $h(z_s)$ of the fluid. The latter can
be defined by \cite{Hill86}:
\begin{equation}
h(z_s) = \lim_{\rho_b\to 0} \frac{\rho_b}{z_b(\rho_b)}
\end{equation}
where $z_b=\Lambda_b^{-3} \exp(\beta \mu_b)$ is the fugacity of species $b$.
It follows that
\begin{equation} \label{Henrylaw}
h(z_s) = \exp(-\beta \omega_1(z_s)).
\end{equation}
$h(z_s)$ does not depend on $b$-$b$ interactions; deviations of $h(z_s)$ from
unity reflect the average effect of $b$-$s$ interactions at a fixed fugacity
$z_s$ of small spheres. Thus, given some means of calculating 
$c_b^{(1)}(\infty)$, in the limit $\rho_b \to 0$, one can determine the 
Henry's law constant.

\subsubsection{Two body term $\Omega_2$}

The two body term $\Omega_2$ is given by:
\begin{equation} \label{pairpot}
\Omega_2(N_b,z_s;\{{\bf r}^b\}) = \sum_{i<j}^{N_b} \omega_2(z_s;|{\bf r}_i^b-
{\bf r}_j^b|),
\end{equation}
where the pair potential $\omega_2$ is defined by \cite{Dijkstra98}
\begin{equation}
\exp(-\beta \omega_2(z_s;r_{ij}^b)=\frac{\langle \exp(-\beta H_{bs}^{(2)})
\rangle_{z_s}}{\langle\exp(-\beta H_{bs}^{(1)})\rangle_{z_s}^2}
\end{equation}
with $r_{ij}^b\equiv |{\bf r}_i^b-{\bf r}_j^b|$ the distance between the 
centers of the big spheres. Equivalently we can use the variable
$h\equiv r_{ij}^b-\sigma_{bb}$, i.e. the distance between the surfaces of two 
big spheres. $\omega_2(z_s;r_{ij}^b)\equiv W(h=r_{ij}^b-\sigma_{bb})$ is the 
grand potential difference between a sea of small spheres, at fugacity
$z_s$, containing two big spheres at finite separation $r_{ij}^b$ and one in
which the separation $r_{ij}^b=\infty$. Although the positions of two big
particles are involved, the big-big interaction does not enter explicitly into
the calculation of $\omega_2$; the diameter $\sigma_{bb}$ merely acts as an
external parameter which restricts the minimum separation to 
$r_{ij}^b=\sigma_{bb}$. In other words, for a given $z_s$, $\sigma_{bs}$ 
and $\sigma_{ss}$, the calculation of $\omega_2$ is the same for both additive
and non-additive systems. 

$\omega_2$ can be identified with the well-known depletion potential between
the two big spheres \cite{Dijkstra98,Roth00} and expressed in terms of the
one-body direct correlation function:
\begin{equation} \label{dep}
\beta \omega_2(z_s;{\bf r}) = \lim_{\mu_b\to -\infty}
(c_b^{(1)}(\infty)-c_b^{(1)}({\bf r}))
\end{equation}
where $c_b^{(1)}({\bf r})$ refers to an inhomogeneous situation in which a big
sphere fixed at the origin exerts its field on the small spheres and a big
(test) particle is inserted at ${\bf r}$ \cite{Roth00}.

\subsubsection{Three body and higher order terms $\Omega_{n\geq 3}$}

The three-body term $\Omega_3$ can be written as a sum of three-body potentials
$\omega_3(z_s;{\bf r}_{i,j,k}^b)$ which can, in turn, be expressed in terms
of ensemble averages $\langle \exp(-\beta H_{bs}^{(n)}) \rangle_{z_s}$ with
$n=1,2,3$ \cite{Dijkstra98}. Once again big-big interactions are not involved
in the calculation and $\sigma_{bb}$ simply specifies the physically allowed
configurations of the three big spheres. Clearly non-additivity plays no role.
The same argument applies for the higher-body ($n>3$) contributions to
$\Omega$, although the specification of the allowed configurations of the big
spheres becomes increasingly complicated as $n$ increases.

In practice the calculation of $\Omega_n$ for $n \geq 3$ is tedious and 
determining the phase behavior for effective Hamiltonians which include these 
and higher-body interactions would be very cumbersome. Contributions to the 
many body terms arise from two mechanisms: (1) Directly, if for a given 
$\sigma_{bb}$, $\sigma_{bs}$ is large enough to allow the overlap of more than
two depletion layers. (2) Indirectly, if correlations between the small 
particles, present for all non-zero $\sigma_{ss}$, induce interactions between 
more than two particles. Contributions from mechanism (1) to $\Omega_n$ with 
$n \geq 3$ are identically zero for 
$(2\sigma_{bs}-\sigma_{bb})/\sigma_{bb} \leq 2/\sqrt{3}-1=0.1547$ 
\cite{Dijkstra99}, while for 
$(2\sigma_{bs}-\sigma_{bb})/\sigma_{bb} \leq \sqrt{3/2}-1= 0.2247$ the 
contributions to $\Omega_n$ with $n \geq 4$ terms are zero\cite{Louis01a} 
etc...  If $\sigma_{ss} > 0$, then mechanism (2) will induce additional 
contributions to the many-body terms at all values of $\sigma_{bs}$ and 
$\sigma_{bb}$.

For the extreme non-additive AO model, where $\sigma_{ss}=0$ but $\sigma_{bs}$ 
is non-zero, only mechanism (1) contributes to the many-body terms.  An 
explicit form for the three-body term can be calculated\cite{Goulding00}, but 
this is still very tedious to evaluate.

For {\em additive} binary hard-sphere mixtures with 
$q=\sigma_{ss}/\sigma_{bb} \leq 0.1$, where only the indirect mechanism (2) 
contributes, three-body contributions seem to be small\cite{Biben96}. Recent 
DFT calculations\cite{Goulding00,Melchionna00} of the three-body potentials 
show that these are still much smaller than the two-body potentials for 
$q=0.2$, where both mechanisms contribute, and that the indirect mechanism (2) 
can have an important effect on their shape.

There is strong evidence from direct simulation studies of the additive binary
system that retaining only two-body contributions in the effective Hamiltonian 
provides a very good account of the equilibrium phase behavior for 
$q \leq 0.2$ \cite{Dijkstra98}. Similar conclusions were reached for a
lattice version of the AO model\cite{Meijer94}.  In this study we shall focus 
on such highly asymmetric systems and neglect three and higher-body 
contributions to the effective Hamiltonian. With this assumption it follows 
that the structure of the homogeneous fluid (equilibrium correlation functions
of the big spheres) is determined solely by the effective pairwise potential
\begin{equation} \label{effpot}
\Phi_{eff}(z_s;r) = \Phi_{bb}(r) + \omega_2(z_s;r),
\end{equation}
since $\Omega_0$ and $\Omega_1$ do not depend on the coordinates of the big
particles \cite{Dijkstra00}. It is also straightforward to show that the 
{\em phase equilibria} of the binary mixture do not depend on the zero and 
one-body term \cite{Dijkstra98}. However these two terms do influence the 
total pressure and compressibility of the mixture \cite{Dijkstra00}.

To complete this sub-section it is necessary to explain how to convert from
$\rho_s^r$, the number density of the small spheres in the reservoir to 
$\rho_s$, the actual value of the small sphere density in the mixture. The 
average number of small spheres in the mixture is given by the thermodynamic
relation
\begin{equation}
\langle N_s \rangle_{z_s} = - \frac{\partial F(N_b,V,z_s)}
{\partial \mu_s} = - \left \langle \frac{\partial \beta \sum_{n=0}^{\infty} 
\Omega_n}{\partial \ln z_s} \right \rangle_{z_s}.
\end{equation}
In Ref.~\cite{Dijkstra98} it was shown that an accurate approximation for the 
conversion can be obtained for additive hard-spheres with high asymmetry,
$q\leq 0.1$, by truncating the expansion after the one-body term so that the 
average number of small spheres can be evaluated approximately from the formula
\begin{equation} \label{convert}
\langle N_s \rangle_{z_s} \approx - \left(\frac{\partial \beta (\Omega_0+
\Omega_1)}{\partial \ln z_s}\right)_{N_b,V}.
\end{equation}
The required density $\rho_s = \langle N_s \rangle_{z_s}/V$. We shall re-visit
this approximation in the next subsection.

\subsection{Evaluation of $\Omega_0$, $\Omega_1$, and $\Omega_2$ within DFT}
\label{sec:dft}

\subsubsection{Rosenfeld's fundamental measure DFT}

In order to make the mapping presented in Sec.~\ref{sec:formal} explicit for 
the model of interest, namely the binary mixture of non-additive hard spheres, 
we calculate the terms $\Omega_0$, $\Omega_1$, and $\Omega_2$ within the 
framework of Rosenfeld's fundamental measure DFT \cite{Rosenfeld89} - a theory
constructed for additive hard-sphere mixtures. As mentioned earlier, the
reason why we can apply a theory constructed for {\em additive} binary 
mixtures is that only the interaction potentials $\Phi_{ss}$, between two small
spheres, and $\Phi_{bs}$, between a big and a small sphere, enter into the 
calculation of $\Omega_n$. This argument substantiates further the intuitive
picture presented in Ref.~\cite{Roth01}.

What is required for calculating $\Omega_1$ and $\Omega_2$ is some means of
determining $c_b^{(1)}$ in the limit of vanishing density of big spheres. DFT
provides a suitable route since \cite{Evans79}
\begin{equation}
c_b^{(1)}({\bf r}) = - \beta \frac{\delta {\cal F}_{ex}[\rho_b,\rho_s]}
{\delta \rho_b({\bf r})}
\end{equation}
where ${\cal F}_{ex}[\rho_b,\rho_s]$ is the excess (over ideal) intrinsic
Helmholtz free energy functional of the mixture \cite{Roth00}. Thus given some
prescription for the mixture functional one can calculate all the necessary
ingredients. Rosenfeld's fundamental measure theory \cite{Rosenfeld89} 
supplies an approximate functional ${\cal F}_{ex}$ for an additive mixture of 
hard spheres of the form
\begin{equation}
\beta {\cal F}_{ex}[\rho_b({\bf r}),\rho_s({\bf r})]=\int d^3 r 
\Psi(\{n_\alpha\}),
\end{equation}
with weighted densities $n_\alpha$ which depend on the fundamental geometrical
measures of the spheres constituting the mixture. There are four scalar and 
two vector weight functions $\mbox{w}_\alpha^i$, with $1 \leq \alpha \leq 6$. 
Details can be found in Ref.~\cite{Rosenfeld89}; the weights depend on the 
radii $R_i$ of each species. For a binary mixture the weighted densities are
\begin{equation}
n_\alpha({\bf r}) = \sum_{i=s,b} \int d^3 r' \rho_i({\bf r}') 
\mbox{w}_\alpha^i({\bf r}-{\bf r}').
\end{equation}
It is important to realize that once the reduced free energy density 
$\Psi(\{n_\alpha\})$ is specified the mapping described in 
Sec.~\ref{sec:formal} is completely determined within this (approximate) 
DFT framework. We choose to apply the {\em original} Rosenfeld functional 
\cite{Rosenfeld89} 
\begin{equation}
\Psi(\{n_\alpha\}) = -n_0 \ln(1-n_3) + 
\frac{n_1 n_2 - \vec{n}_1 \cdot \vec{n}_2}{1-n_3}+
\frac{n_2^3 - 3 n_2 \vec{n}_2 \cdot \vec{n}_2}{24 \pi (1-n_3)^2}.
\end{equation}
Recall that the two-body direct correlation functions $c^{(2)}_{ij}$, with
$i$, $j$ $\in$ $b$, $s$, obtained by taking two functional derivatives of this
functional reduce to the Percus-Yevick 
$\mbox{$c^{(2)}_{ij}(|{\bf r}-{\bf r}'|)$}$ for
a homogeneous hard-sphere mixture. The Rosenfeld functional has proven to
be extremely successful in describing the structure of the inhomogeneous fluid
phases of hard-sphere mixtures. If solid phases are to be considered, 
modifications to the original Rosenfeld functional should be made 
\cite{Rosenfeld96}.

\subsubsection{Calculating $ \Omega_0$ from DFT}

We begin the explicit mapping by noting that the equation of state underlying
the Rosenfeld functional \cite{Rosenfeld89} is $p_{PY}^c$, the Percus-Yevick 
compressibility equation of state for additive hard-sphere mixtures. In order 
to calculate the zero-body term, however, we need only the equation of state 
for a one-component fluid so we set
\begin{equation} \label{eos}
\beta p_s(z_s) = \beta p_{PY}^c(z_s) \equiv \rho_s^r 
\frac{1+\eta_s^r+(\eta_s^r)^2}{(1-\eta_s^r)^3},
\end{equation}
with $\eta_s^r\equiv \pi \sigma_{ss}^3 \rho_s^r/6$, the reservoir packing 
fraction of the small spheres.  $\Omega_0$ follows directly from 
Eq.~\ref{Omega0}.

\subsubsection{Calculating $\Omega_1$ from DFT}

Using the Rosenfeld functional the direct one-body correlation function
$c_b^{(1)}$ can be written as \cite{Roth00}
\begin{equation} \label{cb1}
c_b^{(1)}({\bf r}) = -\sum_\alpha \int d^3 r' \left( 
\frac{\partial \Psi(\{n_\alpha\})}{\partial n_\alpha}\right)_{{\bf r}'}~ 
\mbox{w}_\alpha^b({\bf r}-{\bf r}').
\end{equation}
In the dilute limit in which the density of the big spheres $\to 0$, the 
weighted densities depend only on the density profile $\rho_s({\bf r})$ of the
small spheres:
\begin{equation} \label{ndilute}
n_\alpha^{dilute}({\bf r}) = \int d^3 r'~ \rho_s({\bf r}') 
\mbox{w}_\alpha^s({\bf r}-{\bf r}').
\end{equation}
Note that $\rho_s({\bf r})$ then corresponds to a one-component fluid of 
species $s$ \cite{Roth00}. In {\em bulk} the density profile of the small 
spheres $\rho_s({\bf r})$ is constant and equal to $\rho_s^r$, the vector
weighted densities $\vec{n}_\alpha$ vanish and scalar weight 
functions reduce to those of a one-component fluid, i.e. 
$n_3\to\eta_s^r$, $n_2\to 6\eta_s^r/\sigma_{ss}$, 
$n_1\to 3\eta_s^r/(\pi\sigma_{ss}^2)$ and 
$n_0 \to 6 \eta_s^r/(\pi \sigma_{ss}^3)$. $c_b^{(1)}(\infty)$ can be evaluated
explicitly and the result expressed as
\begin{equation} \label{cbbulk}
\beta \omega_1(z_s) = \lim_{\mu_b\to -\infty} - c_b^{(1)}(\infty)= 
\frac{4 \pi R_b^3}{3} \beta p_s(z_s)+ 4 \pi R_b^2 \beta \gamma(z_s) + 
\frac{R_b}{\sigma_{ss}} \frac{6 \eta_s^r}{1-\eta_s^r}-\ln(1-\eta_s^r),
\end{equation}
where the first term corresponds to the partial derivative 
$\partial/\partial n_3$, the second to $\partial/\partial n_2$ and so on. As 
the first term is proportional to the pressure of the reservoir of small 
spheres, given by Eq.~(\ref{eos}) and the second term is proportional to the 
planar `surface tension' 
\begin{equation} \label{gamma}
\beta \gamma(z_s) = \frac{3 \eta_s^r (2+\eta_s^r)}{2 \pi \sigma_{ss}^2 
(1-\eta_s^r)^2},
\end{equation}
these have a natural interpretation as $p_s \Delta V$ and $\gamma \Delta A$
terms, respectively. Although it is more difficult to give a physical
interpretation of the last two terms in Eq.~(\ref{cbbulk}) they are important,
for example in obtaining the correct low density limit -- see below.
Since $-\beta^{-1} c_b^{(1)}(\infty)$ is the excess chemical potential of 
species $b$ in a uniform (bulk) mixture we can also determine this quantity 
starting from the bulk excess free energy density, differentiating w.r.t. the 
bulk density $\rho_b$ and then taking the limit $\rho_b\to 0$. The result is 
identical to that in Eq.~(\ref{cbbulk}).

Although we have derived this result starting from a theory developed for an 
additive mixture once we have a taken the limit $\rho_b\to 0$ the big-big 
interaction is not relevant. It follows that for an arbitrary non-additive 
mixture, $R_b$ entering Eq.~(\ref{cbbulk}) should be defined as 
$R_b\equiv \sigma_{bs}-\sigma_{ss}/2$. In the particular case of an additive 
mixture $R_b$ reduces to $\sigma_{bb}/2$.

Note that the {\em form} of Eq.~(\ref{cbbulk}) is the same as that given by 
Henderson \cite{Henderson83} in a scaled particle theory for
$c_b^{(1)}(\infty)$. However, in Henderson's treatment $c_b^{(1)}(\infty)$ is 
expressed in powers of the variable $R\equiv \sigma_{bs}$. If one converts his 
Eq.~(54) to an expression in terms of $R_b = R-\sigma_{ss}/2$ one recovers 
{\em precisely} Eq.~(\ref{cbbulk}). In other words, his scaled particle 
analysis agrees completely with the present DFT approach, attesting further to
the consistency of the latter. Note also that in the low density limit
$\rho_s^r\to 0$, Eq.~(\ref{cbbulk}) implies 
$\beta \omega_1(z_s)\to 4\pi \sigma_{bs}^3 \rho_s^r/3$ which is the 
{\em exact} limiting value -- see Appendix.

\subsubsection{Converting from the reservoir packing fraction $\eta_s^r$ to 
the packing fraction $\eta_s$ in the mixture}

\begin{figure}
\centering\epsfig{file=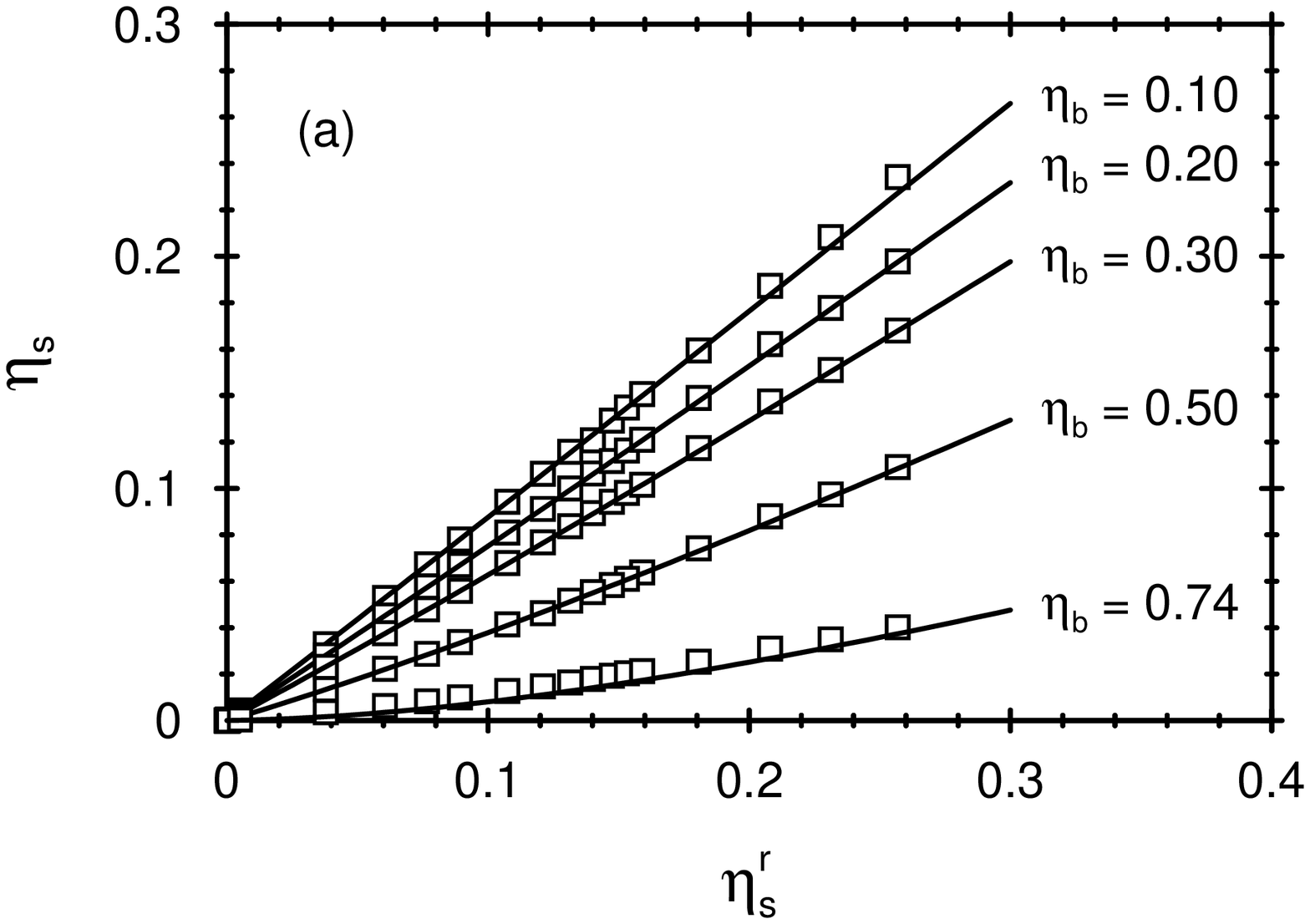,width=12cm}
\centering\epsfig{file=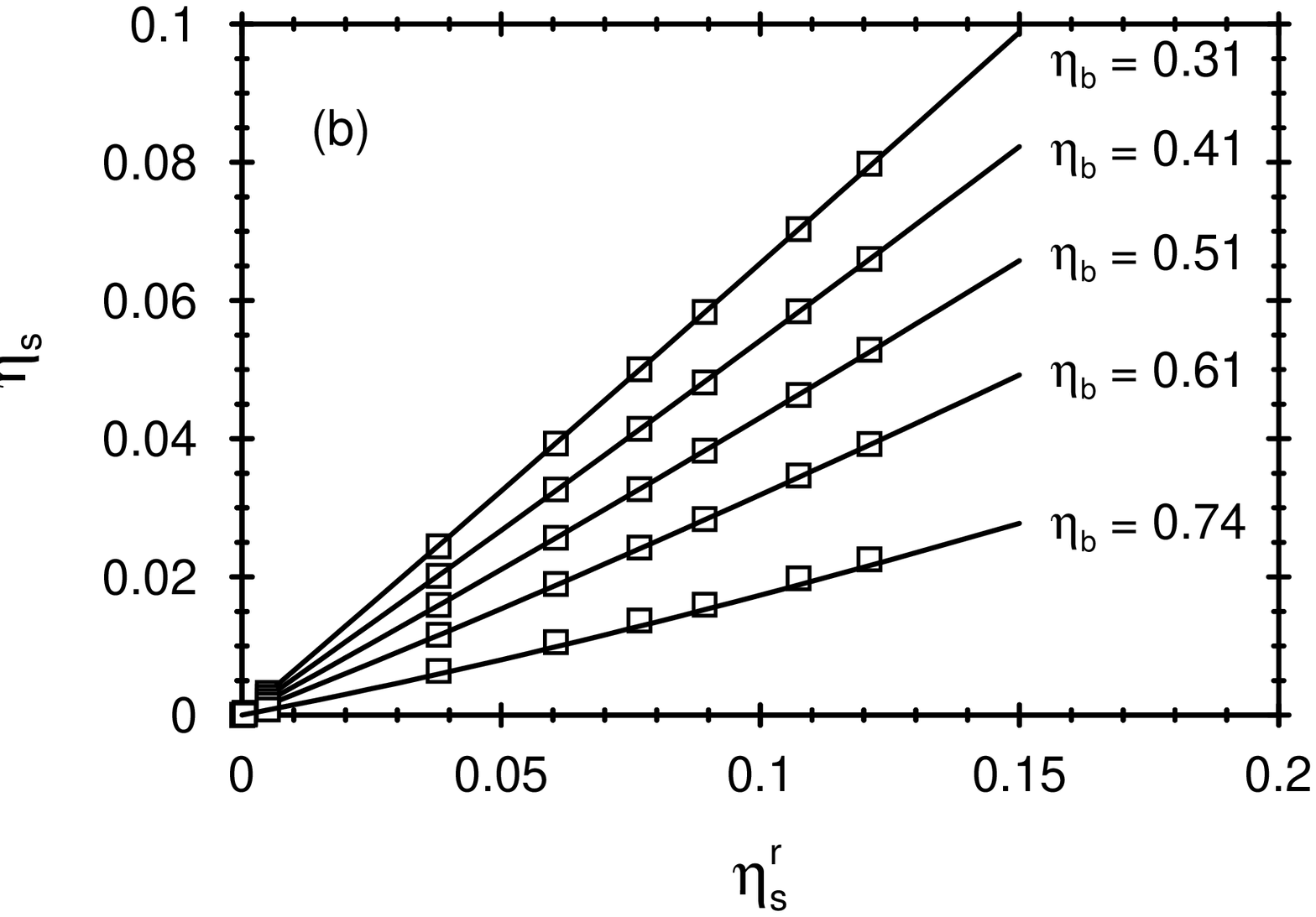,width=12cm}
\vspace{0.5cm}
\caption{\label{fig:convert} The small sphere packing fraction $\eta_s$ of an
additive binary hard-sphere mixture, with size ratio (a) $0.10$ and (b)
$0.05$, versus that of the reservoir $\eta_s^r$ for several big sphere packing
fractions $\eta_b$. The squares denote the direct simulation data of
\protect\cite{Dijkstra98} while the lines denote the results of 
Eq.~(\ref{etas}). Note the significant deviations from linearity in both theory
and simulations.}
\end{figure}

Given explicit formulae for $\Omega_0$ and $\Omega_1$ we can employ
Eq.~(\ref{convert}) to obtain an explicit conversion between $\eta_s$, the 
packing fraction of the small spheres in the system, and $\eta_s^r$, the 
packing fraction in the reservoir. To this end we first determine the fugacity
$z_s=\rho_s^r \exp(\beta \mu_s^{ex})$ of the small spheres within the 
framework of the Rosenfeld DFT approach. $\mu_s^{ex}$, the excess chemical 
potential of the pure small sphere fluid, is given by the Percus-Yevick 
(compressibility) result so that
\begin{equation} \label{fugacity}
z_s = \frac{6 \eta_s^r}{\pi \sigma_{ss}^3 (1-\eta_s^r)}
\exp \left(\frac{14 \eta_s^r-13(\eta_s^r)^2+5 (\eta_s^r)^3}{2(1-\eta_s^r)^3}
\right).
\end{equation}
Using the expressions for $\Omega_0$ and $\Omega_1$ given in this section
together with Eqs.~(\ref{convert}) and (\ref{fugacity}) we obtain for the 
packing fraction of small spheres in the system 
$\eta_s = \pi \sigma_{ss}^3 \rho_s/6$:
\begin{equation} \label{etas}
\eta_s = (1-\eta_b) \eta_s^r - 3 q_{eff}~ \eta_b \eta_s^r 
\frac{1-\eta_s^r}{1+2\eta_s^r} - 3 q_{eff}^2~ \eta_b \eta_s^r
\frac{(1-\eta_s^r)^2}{(1+2 \eta_s^r)^2} - q_{eff}^3~ \eta_b \eta_s^r
\frac{(1-\eta_s^r)^3}{(1+2 \eta_s^r)^2},
\end{equation}
where $q_{eff} \equiv \sigma_{ss}/2 R_b$ is the effective size ratio. In the 
limit $\eta_s^r\to 0$ this result reduces to
\begin{equation}
\eta_s/\eta_s^r = (1-\eta_b(1+q_{eff})^3)
\end{equation}
which is the standard excluded volume expression, appropriate to an ideal gas 
of small particles \cite{Dijkstra98}. However, for non-zero densities of small
spheres Eq.~(\ref{etas}) predicts a non-linear dependence of $\eta_s$ on 
$\eta_s^r$. This is illustrated in Fig.~\ref{fig:convert} where we plot 
$\eta_s$ versus $\eta_s^r$ for {\em additive} hard-sphere mixtures with (a) 
$q=0.1$ and (b) $q=0.05$. Our results are compared with those of direct 
Monte-Carlo simulations of the binary mixture \cite{Dijkstra98}. The agreement 
between theory and simulation is excellent, with a small deviation occurring 
at $\eta_b=0.74$ and $0.10$ for $q=0.1$. Note that the free-volume theory of
Lekkerkerker and Stroobants \cite{Lekkerkerker93}, which asserts that
$\eta_s/\eta_s^r = \alpha(z_s=0; \eta_b)$, where $\alpha(z_s=0;\eta_b)$ is the
free volume fraction evaluated for zero fugacity of the small spheres,
significantly underestimates $\eta_s/\eta_s^r$ at higher values of $\eta_s^r$
\cite{Dijkstra98}. We conclude that retaining only the two leading terms
$\Omega_0$ and $\Omega_1$ and employing PY theory for these quantities provides
an accurate approximation for the free volume fraction $\eta_s/\eta_s^r$, at
least for small values of $q$.

For completeness we should mention that the theory employed in 
Ref.~\cite{Dijkstra98} to calculate $\eta_s/\eta_s^r$ used the 
Carnahan-Starling result for $z_s$ and Henderson's expression for 
$c_b^{(1)}(\infty)$ but with an empirical modification of the leading ($R_b^3$)
term. It is now clear that there was no need to make such a modification; the
confusion arose from the improper identification of the parameter $R$ in
Henderson's theory. Fortunately the numerical results presented in Fig.~13 of
Ref.~\cite{Dijkstra98} are very close to those given by the present, fully
consistent theory. For future applications we recommend that 
Eq.~(\ref{etas}) should be used.

\subsubsection{Calculating $\Omega_2$ from DFT} 

The two-body contribution $\Omega_2$, given in Eq.~(\ref{pairpot}), requires
the calculation of the depletion potential $\omega_2(z_s;r)$ given by 
Eq.~(\ref{dep}). This can be carried out using the procedure described in
Ref.~\cite{Roth00}. We first calculate the (inhomogeneous) equilibrium density
profile $\rho_s(r)$ of the small spheres near a fixed big sphere of radius 
$R_b$. This is used in Eq.~(\ref{ndilute}) to determine the relevant weighted
densities which then determine $c_b^{(1)}(r)$ via Eq.~(\ref{cb1}). 

We emphasize that the mapping of a depletion potential in a non-additive 
system onto one in an additive mixture is exact and has been applied in a
recent study of generalized effective potentials \cite{Louis01}. 

\section{Applications} \label{sec:application}

\subsection{Effect of non-additivity on the shape of depletion potentials}
\label{sec:deppot}

\subsubsection{Changes of type {\bf (a)}: $\sigma_{ss}$ varies, but 
$\sigma_{bs}$ is fixed}

Following the procedure of \cite{Roth01} we follow route {\bf (a)} and 
demonstrate that for a given (fixed) size ratio $q$ and packing fraction 
$\eta_s^{add}$ of the additive mixture the effects of non-additivity $\Delta$ 
on the shape of the depletion potential  are very strong. 

\begin{figure}
\centering\epsfig{file=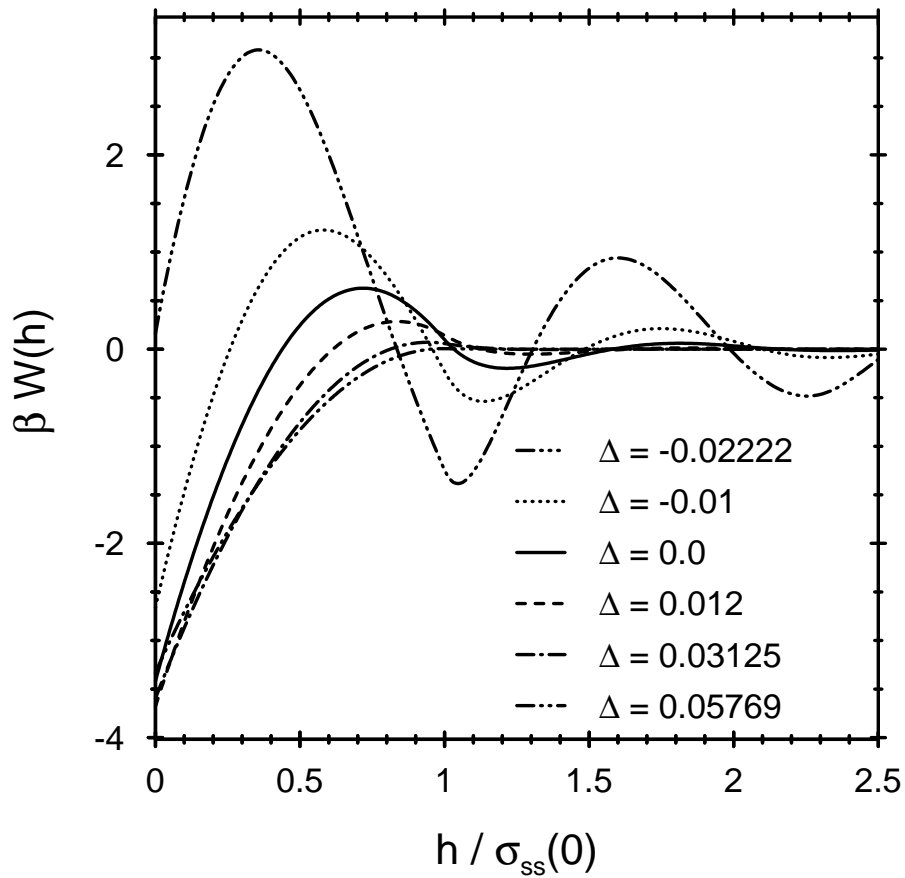,width=12cm}
\vspace{0.5cm}
\caption{\label{fig:pot1} The depletion potential 
$W(h)\equiv \omega_2(z_s;\sigma_{bb}+h)$
between two big hard spheres in a sea of small hard-spheres calculated for a
size ratio $q=0.1$ and a range of non-additivities $\Delta$ treated according
to route {\bf (a)}. The number density of the small spheres, $\rho_s^r$, is 
fixed with packing fraction 
$\eta_s^{add}\equiv \rho_s^r \pi (\sigma_{ss}^{add})^3/6 = 0.2$. $\Delta=0$
corresponds to an additive hard-sphere mixture. $h$ is the separation between
the surfaces of the big spheres and $\sigma_{ss}(0)=q \sigma_{bb}$.}
\end{figure}

Choosing a size ratio $q=0.1$ and a fixed packing fraction in the additive 
mixture of $\eta_s^{add}=0.2$, we can vary the non-additivity parameter 
$\Delta$ between $\Delta=-0.031$, corresponding to the packing fraction 
$\eta_s^r(\Delta)$ of the small spheres, given by Eq.~\ref{eta}, reaching the 
freezing packing fraction $\eta^{freez}=0.494$, and $\Delta=\Delta^{AO}=q$, in
which case $\sigma_{ss} \equiv 0$ and hence $\eta_s^r(\Delta^{AO})\equiv 0$. 
Results for the depletion potentials are given in Fig.~\ref{fig:pot1}. For 
$q=0.1$ and $\eta_s^{add}=0.2$ the depletion potential obtained from DFT for 
additive hard spheres ($\Delta=0$) is in excellent agreement with the results 
of computer simulations \cite{Biben96} -- see the comparison in 
Ref.~\cite{Roth00}. Moreover for $\Delta=\Delta^{AO}=0.1$ we find that our 
calculated depletion potential is indistinguishable from the analytic AO 
result \cite{Roth01}.

For small degrees of positive non-additivity the main effects observed in the 
depletion potential are a weakening of the first repulsive potential barrier, 
due to a decreased packing fraction [given by Eq.~(\ref{eta})], and an increase
in the range of attraction, i.e. the maximum of the potential shifts to larger
separations $h$. Both effects tend to increase the net attraction and this is
reflected in the second virial coefficient, as we shall see later.

\begin{figure}
\centering\epsfig{file=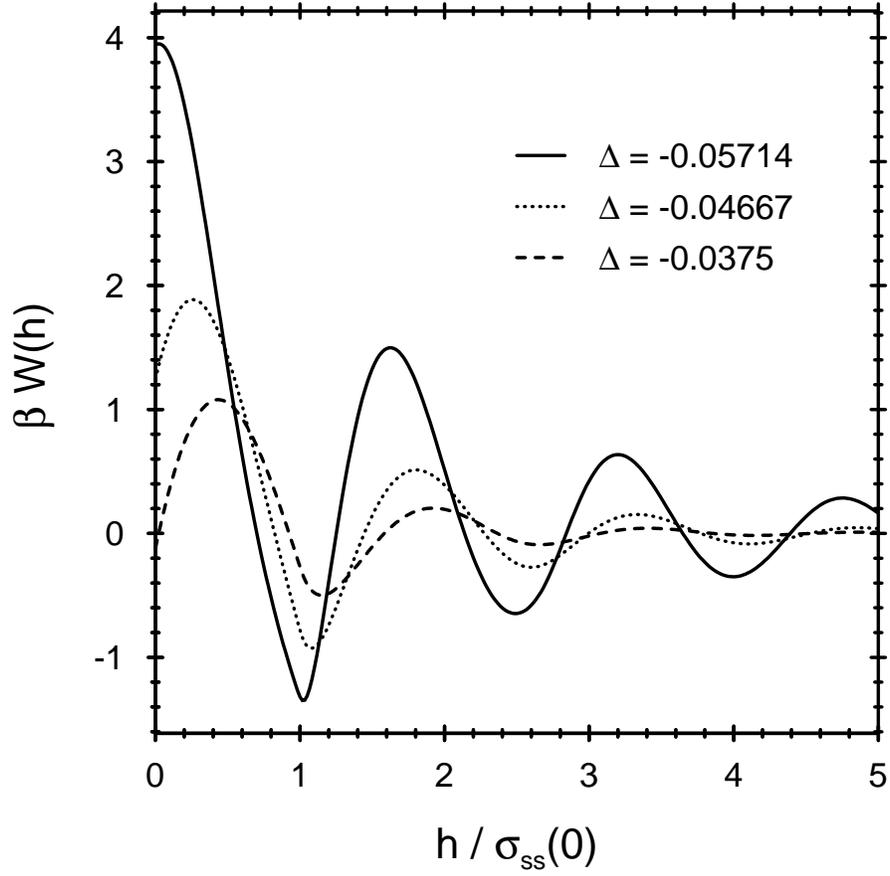,width=12cm}
\vspace{0.5cm}
\caption{\label{fig:pot2} As in Fig.~\ref{fig:pot1} but now the packing 
fraction $\eta_s^{add}$ is fixed at $0.1$. For these negative values of the 
non-additivity $\Delta$ the depletion potential is repulsive near contact and
for $\Delta=-0.05714$ the depletion force is repulsive near contact.}
\end{figure}

For negative values of $\Delta$ the packing fraction $\eta_s^r(\Delta)$ of the
small spheres increases rapidly and the contact value of the depletion 
potential increases sharply -- see Fig.~\ref{fig:pot1}. For sufficiently 
negative values of $\Delta$ the depletion potential can become positive at 
contact while the force near contact remains attractive. If the density 
$\rho_s^r$ of the small spheres is small enough to permit a high degree of 
negative  non-additivity the depletion force near contact can even become 
repulsive. This is illustrated in Fig.~\ref{fig:pot2} for parameters 
$\eta_s^{add}=0.1$ and $q=0.1$. Now $\Delta$ can take values as low as 
$-0.060$ whilst $\eta_s^r(\Delta)$ remains smaller than $\eta_s^{freez}$. 
Depletion potentials for $q=0.1$ and $\eta_s^{add}=0.3$ were presented in 
Fig.~3 of Ref.~\cite{Roth01} and for $q=0.2$ and $\eta_s^{add}=0.1$ were
presented in Fig.~2 of Ref.~\cite{Louis01}; these display similar trends with $\Delta$ as those shown 
here.

\subsubsection{Changes of type {\bf (b)}: $\sigma_{bs}$ varies, but 
$\sigma_{ss}$ is fixed}

\begin{figure}
\centering\epsfig{file=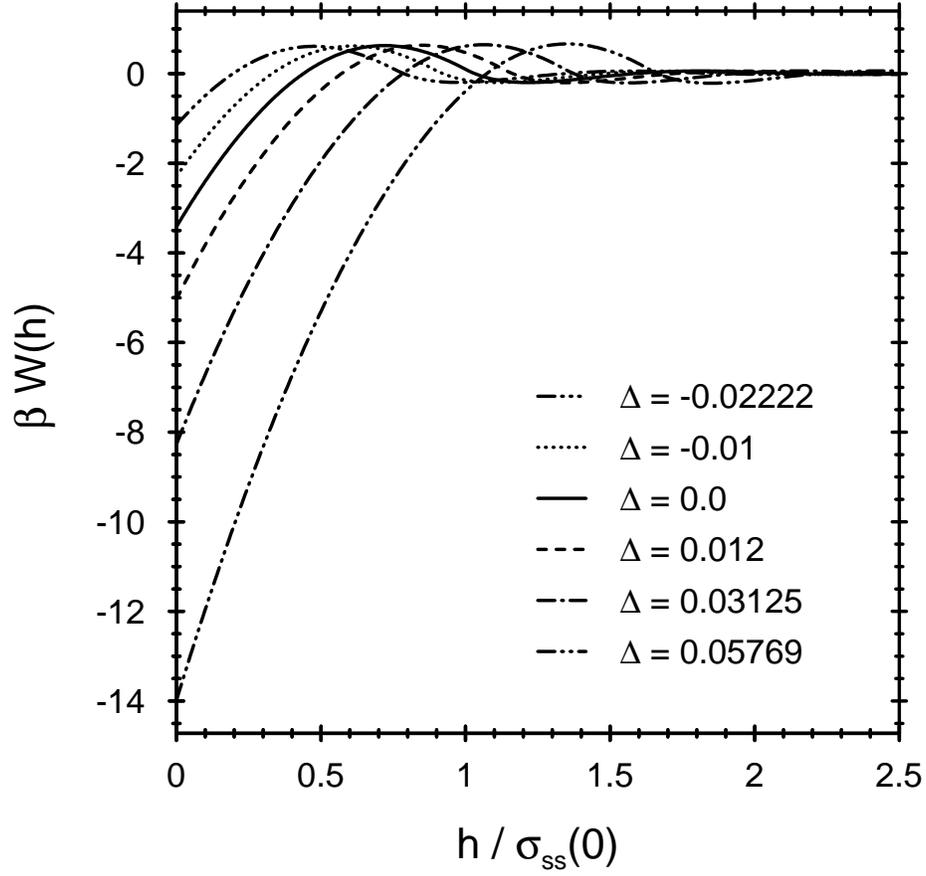,width=12cm}
\vspace{0.5cm}
\caption{\label{fig:pot4} The depletion potential 
$W(h)\equiv \omega_2(z_s;\sigma_{bb}+h)$
between two big hard spheres in a sea of small hard-spheres calculated for a
size ratio $q=0.1$ and a range of non-additivities $\Delta$ treated according
to route {\bf (b)}. The packing fraction in the reservoir $\eta_s^r=0.2$ 
remains constant for all values of $\Delta$. $\Delta=0$ corresponds to an 
additive hard-sphere mixture. $h$ is the separation between the surfaces of 
the big spheres and $\sigma_{ss}(0)=q \sigma_{bb}$. These results should be
contrasted with those in Fig.~\ref{fig:pot1} -- note the difference between
the vertical scales.}
\end{figure}

In Fig.~\ref{fig:pot4} we show the effect of changing $\Delta$ according to
route {\bf (b)}. Now the packing fraction in the reservoir $\eta_s^r$ is fixed
at $0.2$ for all values of $\Delta$ and $\sigma_{bs}$ varies. The results are
very different from those in Fig.~\ref{fig:pot1} which correspond to the
same size ratio, $q=0.1$, and the same $\eta_s^{add}$. In the present case 
increasing $\Delta$ shifts, almost rigidly, the depletion potential to larger 
separations $h$ leading to much deeper and longer ranged attractive wells than 
for $\Delta=0$. Making $\Delta$ increasingly more negative corresponds to
shifting the potential to smaller separations thereby reducing the attraction
and the height of the potential barrier. Depletion potentials for $q=0.2$,
$\eta_s^r=0.2$ were presented in Fig.~3 of Ref.~\cite{Louis01}; these display 
similar trends with $\Delta$ as those shown here.

\subsection{Effect of non-additivity on the second virial coefficient}
\label{sec:b2}

\subsubsection{Changes in $B_2$ as a function of $\Delta$}

In the subsection~\ref{sec:deppot} we demonstrated that introducing a rather 
small degree of either positive or negative non-additivity has a profound 
effect on the shape of the depletion potential. Here we investigate the effect
of non-additivity on the second virial coefficient $B_2$ which measures the net
attraction between two big particles in the sea of small ones. $B_2$ 
corresponds to the total effective pair potential $\Phi_{eff}(z_s;r)$ defined 
in Eq.~(\ref{effpot}). It follows that
\begin{equation} \label{B2}
B_2 = B_2^{HS} + 2 \pi \int_{\sigma_{bb}}^\infty d r 
r^2 \left\{1-\exp\left[-\beta \omega_2 (z_s;r)\right]\right\}
\end{equation}
with $B_2^{HS}=2 \pi \sigma_{bb}^3/3$, the second virial coefficient of the 
pure hard-sphere system. If the depletion potential $\omega_2(z_s;r)$ generates
enough attraction between the two big spheres $B_2$ can become negative 
\cite{deHek81}. Note that $B_2$ is a function of $\Delta$ and $z_s$.

\begin{figure}
\centering\epsfig{file=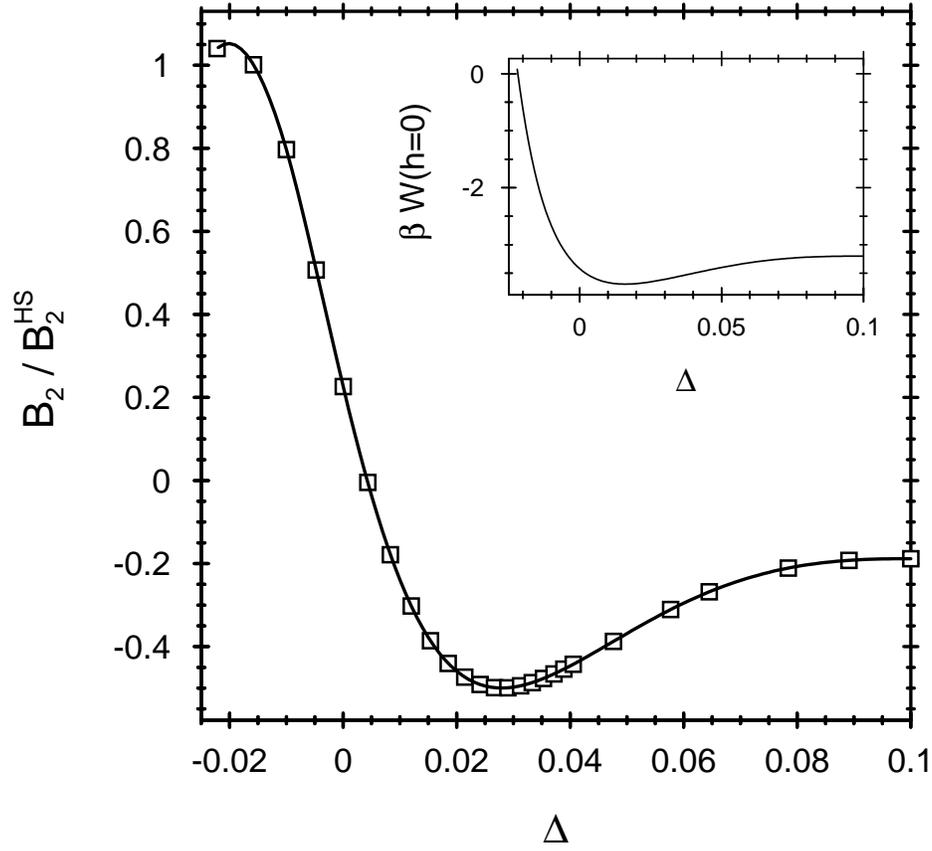,width=12cm}
\vspace{0.5cm}
\caption{\label{fig:b2aolimit} The second virial coefficient of the big 
spheres $B_2$ in units of $B_2^{HS}\equiv 2 \pi \sigma_{bb}^3/3$ as a function
of the non-additivity $\Delta$. These results correspond to the depletion 
potentials of Fig.~\ref{fig:pot1}, i.e. $\eta_s^{add}=0.2$ and $q=0.1$, for 
which the maximum non-additivity is $\Delta^{AO}=0.1$. The inset shows the 
contact value $\beta W(h=0) \equiv \beta \omega_2(z_s;\sigma_{bb})$ of the 
depletion potential as a function of $\Delta$.}
\end{figure}

In Fig.~\ref{fig:b2aolimit} $B_2$ is plotted as a function of $\Delta$ for
$\eta_s^{add}=0.2$ and $q=0.1$. For an intermediate value of 
$\Delta_{min} \approx 0.0279$ we find that $B_2$ takes its minimum value of 
$B_2(\Delta_{min})/B_2^{HS}\approx-0.5$. The variation of $B_2$ with $\Delta$ 
is similar to that of $W(h=0)=\omega_2(z_s;\sigma_{bb})$, the contact value of
the depletion potential, although the latter has its minimum at a slightly 
lower value $\Delta \approx 0.016$ -- see inset of Fig.~\ref{fig:b2aolimit}. 
Provided $\eta_s^r$ is sufficientlty high to generate significant packing
effects the presence of a minimum in $B_2$ in the range 
$0>\Delta_{min}>\Delta^{AO}$ is easily understood \cite{Roth01}. For a given 
size ratio $q$ and density of small particles $\rho_s^r>0$ the Asakura-Oosawa 
depletion potential given by
\begin{equation} \label{AOpot}
W(h) \equiv W^{AO}(h) = - p_{id}(z_s) \Delta V(h),
\end{equation}
where $p_{id}(z_s)=\rho_s^r k_B T$ is the ideal gas pressure and $\Delta V(h)$
is the overlap volume excluded to the centers of the small spheres. $h$
denotes the separation between the surfaces of the two big spheres so that
$\Delta V(h) = 0$ for $h>\sigma_{ss}(0) = q \sigma_{bb}$. $W^{AO}$ is purely 
attractive and should always generates more net attraction than the depletion 
potential of an additive hard-sphere mixture -- provided $\eta_s^r$ is large 
enough that packing effects become significant. Then the depletion potential 
consists of an attractive part close to contact and an oscillatory tail for 
larger separation. Packing effects of the small spheres reduce the range of 
the initial attractive part of the hard-sphere depletion potential compared to
the Asakura-Oosawa potential and for the same value of $q$ and $\rho_s^r$ we 
find $B_2^{AO}<B_2^{add}$, at least for the parameters we studied. Very close 
to the Asakura-Oosawa limit, i.e. $\Delta \lesssim \Delta^{AO}$, where the 
packing fraction of the small spheres, Eq.~(\ref{eta}), is small but non-zero, 
packing effects are minor and the depletion potential is still determined by 
excluded volume considerations. For a non-zero packing fraction, however, the 
pressure of the small sphere fluid is higher than in an ideal gas so that to 
first order in $\eta_s^r$, the virial expansion of this pressure yields
\begin{equation} \label{AO1pot}
W^{AO_1}(h) = (1+4 \eta_s^r(\Delta)) W^{AO}(h).
\end{equation}
In Ref.~\cite{Roth01} we showed that this modified AO approximation is very
accurate for $\eta_s^r(\Delta)<0.01$. According to this approximation the 
depletion potential, Eq.~(\ref{AO1pot}), is more attractive than in the 
Asakura-Oosawa limit so that we can conclude that for an intermediate value of
$\Delta$ the second virial coefficient $B_2$ must have a minimum. It is, 
however, surprising and striking that this minimum is found to be deep,
$(B_2^{HS}-B_2(\Delta_{min}))/(B_2^{HS}-B_2(\Delta^{AO}))\approx 1.26$, and 
located at a rather low degree of non-additivity.

In the additive limit $B_2$ is already positive for this particular mixture 
and a very small degree of negative non-additivity is sufficient to make $B_2$ 
strongly positive. 

\subsubsection{Changes in $B_2$ as a function of $\eta_s^r$ and a criterion
for fluid-fluid phase separation}

In an experimental situation $\Delta$ is not an easily controllable or tunable
parameter and therefore it is most interesting from an experimental point of 
view to consider a fixed value of $\Delta$ and investigate the depletion 
potential and $B_2$ when the reservoir density of the small spheres $\rho_s^r$,
a quantity that can be controlled easily in an experiment, is changed.

If one were to take any real (asymmetric) binary mixture of hard-sphere like
colloidal particles and have some means of determining the three interparticle
pairwise potentials one could, in principle, assign three effective hard-sphere
diameters $\sigma_{bb}$, $\sigma_{bs}$ and $\sigma_{ss}$ using standard liquid
state theories \cite{Louis01}. In general one would not expect these diameters 
to be perfectly additive although the magnitude and sign of the non-additivity
might be difficult to ascertain by any direct measurement. In this subsection 
we demonstrate that a very small degree of non-additivity reveals itself very
clearly in the dependence of $B_2$ on the packing fraction of the small
spheres $\eta_s^r$.

\begin{figure}
\centering\epsfig{file=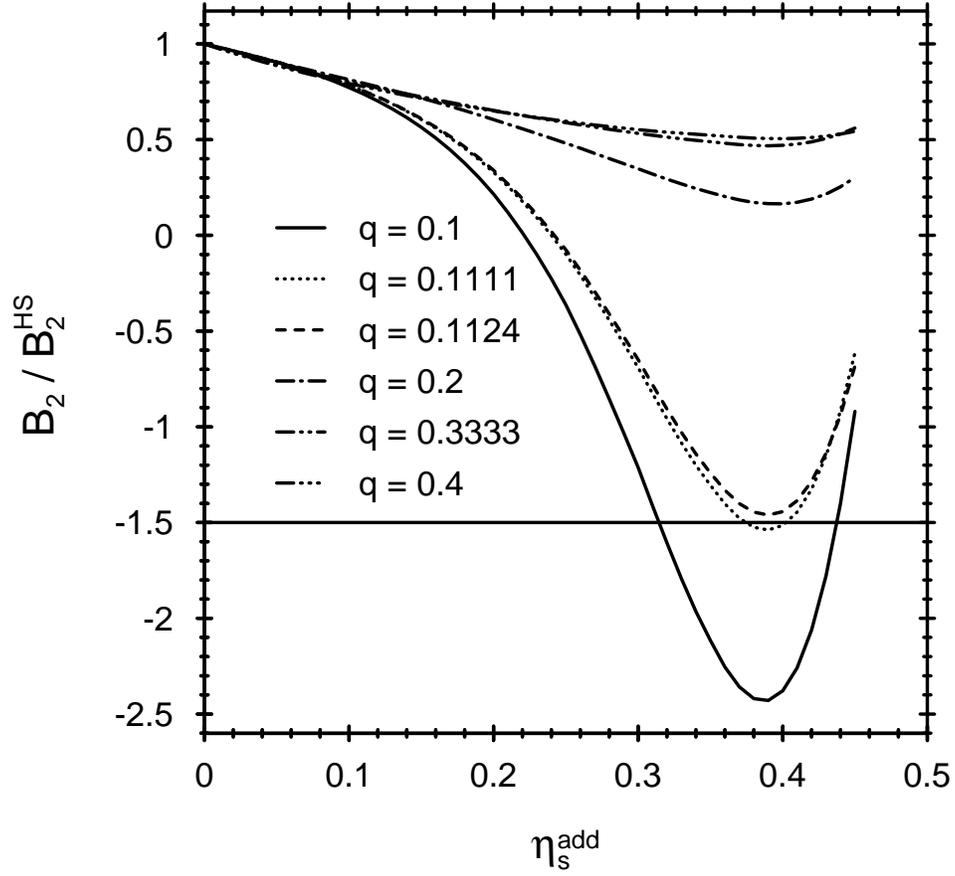,width=12cm}
\vspace{0.5cm}
\caption{\label{fig:b2add} The reduced second virial coefficient of the big 
spheres in {\em additive} hard-sphere mixtures ($\Delta=0$) for various size
ratios $q$ versus $\eta_s^{add}$, the packing fraction of small spheres in the
reservoir. According to the criterion of Ref.~\protect\cite{Vliegenthart00},
(metastable) fluid-fluid phase separation can only occur if $B_2/B_2^{HS}<-1.5$
(horizontal line).}
\end{figure}

To this end we start with an additive mixture and plot in Fig.~\ref{fig:b2add}
$B_2$ expressed in units of the second virial coefficient of a pure 
hard-sphere system $B_2^{HS}$ as a function of $\eta_s^r\equiv \eta_s^{add}$ 
for various size ratios $q$. The qualitative behavior of $B_2$ is the same for
all size ratios: for small values of $\eta_s^r$ the reduced second virial 
coefficient decreases from unity in approximately linear fashion while for 
high packing fractions, $\eta_s^r >0.3$, the decreasing range of attraction 
and the increasing height of the repulsive potential barrier in the depletion 
potential hinder a further decrease of $B_2$ and we find a minimum of the 
second virial coefficient at roughly $\eta_s^r\lesssim 0.4$ and an increase of
$B_2$ upon further increase of $\eta_s^r$.

The quantitative behavior, however, depends very strongly on the value of $q$.
For large size ratios, $q>0.2$ we find that for all packing fractions
$\eta_s^r$ the depletion potential cannot generate enough net attraction to 
make $B_2$ negative. This observation helps us to understand, in terms of the 
depletion potential, the fact that in additive binary hard-sphere mixtures 
with large size ratios $q$ even metastable fluid-fluid phase separation does 
not occur \cite{Dijkstra98}. 

For smaller size ratios the minimum of $B_2$ becomes more pronounced and $B_2$
takes on negative values for a range of $\eta_s^r$. At a size ratio $1:9$
($q=0.1111$) the minimum value of $B_2$ falls below $-1.5 B_2^{HS}$. By 
analyzing a large series of simulation results for a variety of (one-component)
model fluids, Vliegenthart and Lekkerkerker \cite{Vliegenthart00} have shown
recently that the second virial coefficient evaluated at the gas-liquid
critical temperature $T_c$ takes values which lie in a fairly narrow range
around $-1.5 B_2^{HS}$. For the model fluids considered in 
Ref.~\cite{Vliegenthart00} gas-liquid coexistence can only occur if
$B_2/B_2^{HS} < -1.5$. If we assume that this empirical criterion is applicable
to the effective one-component system described by our pairwise potentials 
$\Phi_{eff}$ it follows that only systems with $B_2/B_2^{HS}$ lying below the
horizontal line in Fig.~\ref{fig:b2add} could exhibit (metastable) 
fluid-fluid coexistence. It is important to emphasize that the criterion is
empirical and that it was developed for model fluids in which the attractive
part of the pairwise potential is monotonically increasing with interparticle
separation $r$, unlike our present effective potentials. Moreover the criterion
does not predict whether the gas-liquid coexistence is stable or metastable 
w.r.t. fluid-solid coexistence. Recall that the shorter range of the attractive
potential the more likely is the gas-liquid transition to become metastable
\cite{Vliegenthart00}. In a simulation study of the effective one-component
Hamiltonian for an additive hard-sphere mixture, metastable fluid-fluid phase 
separation was found for $q=0.1$ and $q=0.05$ \cite{Dijkstra98}. For $q=0.1$ 
fluid-fluid coexistence occurred for $\eta_s^r \gtrsim 0.29$, whereas for 
$q=0.05$ this occurred for $\eta_s^r \gtrsim 0.165$. These results are in 
keeping with the predictions of the empirical criterion. Note that the second 
intersection of $B_2/B_2^{HS}$ with the horizontal line for $q=0.1$ in 
Fig.~\ref{fig:b2add} suggests a possible upper critical point near 
$\eta_s^r=0.43$ \cite{Louis01}. 

If we introduce a very small degree of {\em positive} non-additivity according
to route {\bf (a)}, i.e. we keep $\sigma_{bb}$ and $\sigma_{bs}$ constant so 
that $\sigma_{ss}(\Delta)$ becomes smaller than for the additive case (see
Eq.~\ref{sigma}), we find a dramatically different situation. In order to 
be of relevance to an experimental situation, where it might be impossible to 
rule out a small degree of non-additivity, we set $\Delta=q/20$ for each choice
of $q$.

\begin{figure}
\centering\epsfig{file=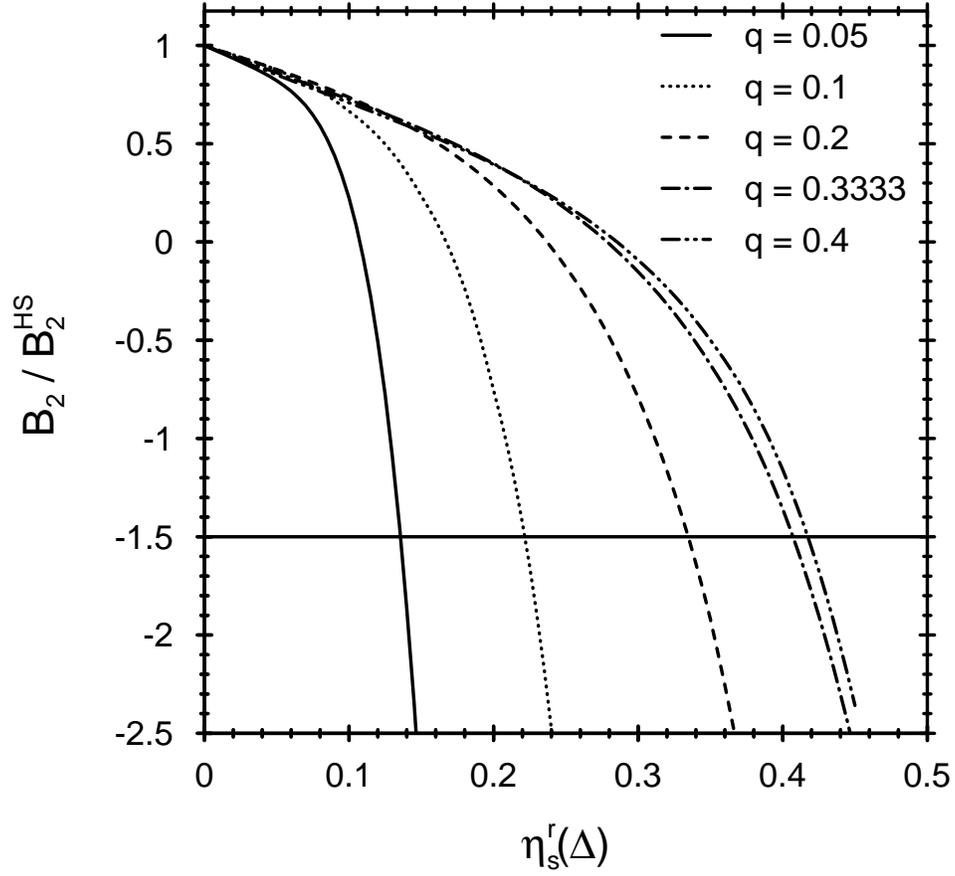,width=12cm}
\vspace{0.5cm}
\caption{\label{fig:b2pos} The reduced second virial coefficient of the big
spheres in hard-sphere mixtures with a small {\em positive} non-additivity
$\Delta=+q/20$ for various size ratios $q$ versus $\eta_s^r(\Delta)$, the
packing fraction of small spheres in the reservoir. In contrast to the additive
case, Fig.~\ref{fig:b2add}, $B_2/B_2^{HS}$ always falls below $-1.5$ indicating
that (metastable) fluid-fluid phase separation could occur for all size ratios
shown here.}
\end{figure}

As expected, positive non-additivity leads to a slight increase in the width of
the depletion layer and therefore {\em more} net attraction than in the 
additive case. The behavior of $B_2$ as a function of $\eta_s^r(\Delta)$, 
however, changes qualitatively as can be seen in Fig.~\ref{fig:b2pos}. For all
size ratios considered here the second virial coefficient becomes negative 
and $B_2/B_2^{HS}$ falls below the $-1.5$ line prior to freezing of the small
particles. The smaller $q$, the smaller the value of $\eta_s^r$ when this line 
is crossed. There is no minimum in $B_2$. Thus according to the empirical 
criterion all the mixtures considered here should exhibit (metastable) 
fluid-fluid coexistence. Once again we cannot say whether this transition is 
stable w.r.t. the fluid-solid transition. 

Our results provide some understanding, in terms of the depletion potential
description, of why only small degrees of positive non-additivity might lead
to fluid-fluid phase separation in asymmetric binary hard-sphere mixtures. We
recall that Biben and Hansen \cite{Biben97} found for $q=0.1$, on the basis
of the Barboy-Gelbart \cite{Barboy79} equation of state, that a value 
$\Delta=0.01$ was sufficient to produce a fluid-fluid transition at a total 
packing fraction $<0.5$. Later Dijkstra \cite{Dijkstra98b} carried out a 
series of Gibbs ensemble Monte-Carlo simulations of the binary mixture for 
$q=0.1$ and varying degrees of positive non-additivity. She found that it was 
possible to have fluid-fluid demixing for a total packing fraction $<0.5$, 
provided $\Delta$ was sufficiently large.

On the other hand, introducing a small degree of {\em negative} non-additivity 
[again via route {\bf (a)}] decreases the width of the depletion layer compared
to the additive case so that the net attraction should also decrease. We
set $\Delta=-q/20$ for each $q$ and find that these small negative values of 
$\Delta$ are sufficient to change significantly the shape of $B_2$ versus 
$\eta_s^r(\Delta)$ from that of the additive case.

\begin{figure}
\centering\epsfig{file=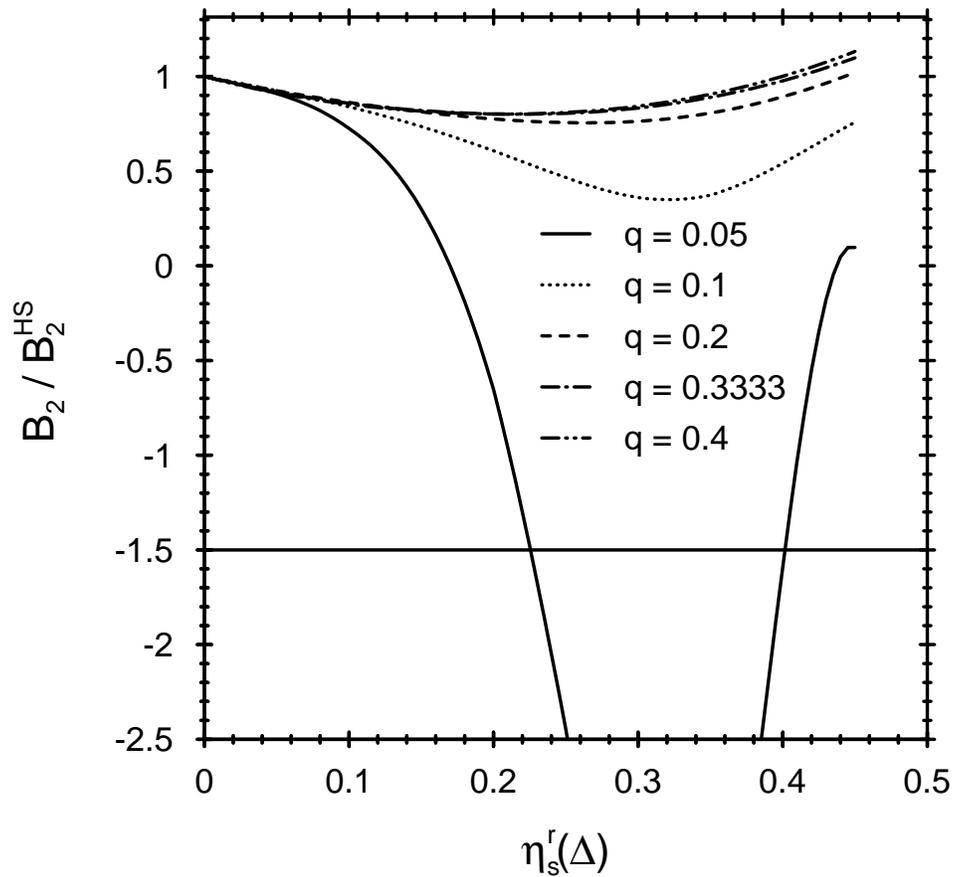,width=12cm}
\vspace{0.5cm}
\caption{\label{fig:b2neg} The reduced second virial coefficient of the big
spheres in hard-sphere mixtures with a small {\em negative} non-additivity
$\Delta=-q/20$ for various size ratios $q$ versus $\eta_s^r(\Delta)$, the
packing fraction of small spheres in the reservoir. Note that $B_2/B_2^{HS}$
falls below $-1.5$ only for the smallest ratio, $q=0.05$, considered here.}
\end{figure}

This is illustrated in Fig.~\ref{fig:b2neg} where we show that for size ratios
$q \gtrsim 0.2$ the second virial coefficient changes little as function of 
$\eta_s^r(\Delta)$. For large values of $\eta_s^r$ we even find that 
$B_2>B_2^{HS}$. For a size ratio of $q=0.1$, which in the additive case was 
sufficiently small for $B_2$ to become strongly negative, $B_2$ remains 
positive for all packing fractions. Only for very small $q$ can the depletion 
potential in non-additive mixtures with negative $\Delta$ generate sufficient 
net attraction to drive $B_2$ negative.

\begin{figure}
\centering\epsfig{file=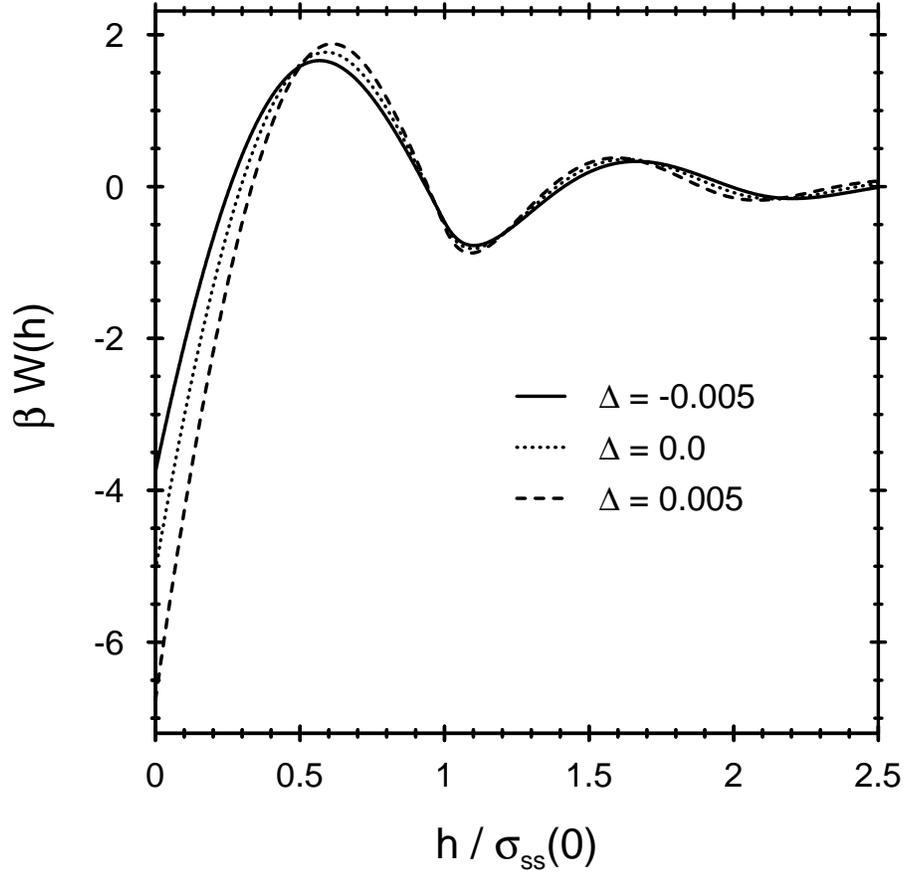,width=12cm}
\vspace{0.5cm}
\caption{\label{fig:pot3} Depletion potentials for $q=0.1$ and fixed 
$\eta_s^r(\Delta)=0.3$ for $\Delta=-0.005$, $0$ and $+0.005$. Although these
potentials do not differ drastically from each other the corresponding $B_2$, 
see Figs.~\ref{fig:b2add}--\ref{fig:b2neg}, take very different values:
$B_2/B_2^{HS}=0.36$, $-1.21$ and $<-9$ for $\Delta=-0.005$, $0$, and $+0.005$,
respectively.}
\end{figure}

We conclude this discussion by emphasizing how sensitive $B_2$ is to changes
in the depletion potential. In Fig.~\ref{fig:pot3} we show the depletion
potentials calculated for $q=0.1$ and fixed $\eta_s^r(\Delta)=0.3$ for three
values of $\Delta$. Although the three potentials appear rather similar they
yield very different values of $B_2$. From 
Figs.~\ref{fig:b2add}--\ref{fig:b2neg} we see that $B_2/B_2^{HS}$ is about
$-1.21$ for $\Delta=0$, is strongly negative ($B_2/B_2^{HS}<-9$) for 
$\Delta=0.005$ (the empirical criterion would predict fluid-fluid phase 
separation) and is positive ($B_2/B_2^{HS}=0.36$) for $\Delta=-0.005$.

\section{Discussion} \label{sec:discussion}

The main results of our study of the equilibrium statistical mechanics of
non-additive binary hard-sphere mixtures may be summarized as follows:

1. The formal technique of integrating out the degrees of freedom of the small
spheres in order to obtain an effective Hamiltonian $H^{eff}$ for the big
spheres is equally valid for non-additive as for additive mixtures, where it
has proved particularly useful \cite{Dijkstra98} for determining the phase
behavior of asymmetric systems. We have provided expressions for zero,
$\Omega_0$, one, $\Omega_1$, and two body contributions to $H^{eff}$ which can
be evaluated in simulations of the small sphere fluid (Sec.~\ref{sec:formal}).

2. We showed that the same contributions are readily calculated using the
fundamental measure DFT of Rosenfeld, a theory developed originally for
additive hard-spheres (Sec.~\ref{sec:dft}). By calculating $\Omega_0$ and
$\Omega_1$ we were able to derive an explicit approximation (\ref{etas}) for
the packing fraction $\eta_s$ of small spheres in the mixture at given
reservoir fraction $\eta_s^r$ and big-sphere fraction $\eta_b$. Comparison
with a previous simulation study for additive mixtures with small size ratios
$q$ showed that the approximation is very accurate (Fig.~\ref{fig:convert}).

3. The two body contribution to $H^{eff}$ is a sum of effective pairwise
potentials $\Phi_{eff}(z_s;r)=\Phi_{bb}(r)+\omega_2(z_s;r)$.
Our DFT approach provides a powerful means of determining the depletion 
potential $\omega_2(z_s;r)$ for non-additive mixtures. Provided three
and higher body terms are small, as is expected for small $q$, it is solely
$\omega_2$ that determines big-big correlation functions and the phase 
equilibria of the mixture.

4. We described two different routes to introducing non-additivity $\Delta$
and gave examples of the depletion potentials and the second virial coefficient
$B_2$ associated with the corresponding effective potential $\Phi_{eff}$
obtained from both routes (Sec.~\ref{sec:application}). What emerges is that
the depletion potential depends strongly on whether route {\bf (a))} 
$\sigma_{bs}$ fixed, $\sigma_{ss}$ varies or route {\bf (b)} $\sigma_{ss}$ 
fixed, $\sigma_{bs}$ varies, is employed. $B_2$ is a sensitive indicator of 
the shape and range of the effective potential and exhibits considerable 
variation with $q$, $\Delta$ and $\eta_s^r$ 
(Figs~\ref{fig:b2aolimit}--\ref{fig:b2neg}).

5. On the basis of the empirical criterion $B_2/B_2^{HS}<-1.5$ 
\cite{Vliegenthart00}, we showed that fluid-fluid phase separation is much
more likely to occur for a small degree of positive non-additivity, $\Delta>0$,
than in additive mixtures, $\Delta=0$, with the same size ratio. Our results
provide a guide to which binary hard-sphere mixtures might exhibit a stable,
w.r.t. fluid-solid, fluid-fluid separation and we hope that these might
stimulate further computer simulation studies.

We conclude by turning to the question of whether the strong effect of 
non-additivity on the depletion potentials and on the virial coefficients 
found in sections \ref{sec:deppot} and \ref{sec:b2} has any implications for 
experiments on colloidal systems.

One-component colloidal suspensions which mimic very closely a hard-sphere 
system can be created because any small residual short-range interactions 
remaining after refractive index matching are very well approximated by hard
spheres with an effective hard-sphere diameter\cite{Pusey86}. But creating a 
truly additive binary colloidal suspension is much more difficult, since this 
implies an additional constraint on the value of the effective hard-sphere 
diameters, namely that $2\sigma_{bs} = \sigma_{bb} + \sigma_{ss}$.  The small 
residual interactions in an experimental system designed to mimic binary 
hard-sphere mixtures can easily introduce non-additivity\cite{Louis01};
non-additivity is probably the rule and perfect additivity the exception. For 
example, in an earlier publication\cite{Louis00}, one of us has shown by a 
simple argument that for both sterically and electrostatically stabilized 
asymmetric binary colloids, $2 \sigma_{bs}$ is likely to be smaller than 
$(\sigma_{bb} + \sigma_{ss})$, which implies a small negative non-additivity. 
This in turn suggests that the well-depth at contact $W(h=0)$ is smaller than 
what would be expected for an additive system. For short-ranged depletion 
systems the location of a fluid-solid liquidus line can be roughly correlated 
to $W(h=0)$\cite{Louis00,Louis01a}; one would therefore expect the experimental
liquidus line to occur at larger values of $\eta_s$ than what is predicted for
a purely additive binary hard-sphere system. Experimental results do seem to 
follow this trend\cite{vanD93}. However, since the experimental phase 
boundaries are typically plotted with the small-particle packing fraction 
$\eta_s$ on the $y$-axis, it is not always clear whether discrepancies with 
the additive theory arise from non-additivity, or from small errors in the 
measurement of $\sigma_{ss}$, which enters $\eta_s$ as $\sigma_{ss}^3$.

Instead of focusing on phase boundaries, we propose that direct measurements 
of the osmotic second virial coefficient $B_2$ as a function of $\eta_s$ 
should be a much more sensitive measure of the existence of non-additivity, 
and may even provide an independent way to ascertain the value of 
$\sigma_{bs}$, which is otherwise very hard to determine. As illustrated in 
Figs~\ref{fig:b2add}, \ref{fig:b2pos} and \ref{fig:b2neg}, different degrees 
of non-additivity induce clear qualitative differences in the dependence of 
$B_2$ on $\eta_s$, implying that one does not require a high level of 
quantitative accuracy in measurements of $B_2$ in order to distinguish clearly
between negative and positive non-additivity.

For colloidal suspensions $B_2$ is typically extracted from the low density 
limit of the osmotic equation of state, measured by static light scattering. 
This is non-trivial since it requires extracting the contribution from the big
particles to the total scattering intensity. Such measurements were first 
carried out by de Hek and Vrij in 1982\cite{deHek82} for a colloid-polymer 
mixture with a size ratio $2 R_g/\sigma_{cc} \approx 1$ (here $R_g$ is the 
radius of gyration of the polymers); they found a clear trend towards negative
values of $B_2$ upon increasing the polymer concentration, which is consistent 
with the expected positive non-additivity in such colloid-polymer systems. In 
contrast, the results in section \ref{sec:b2} imply that for an additive 
binary colloid mixture, size-ratios of $\sigma_{ss}/\sigma_{bb}\lesssim 0.2$ 
are required to drive $B_2$ negative. For negative non-additivity even smaller
size-ratios are required. Such qualitative effects should be visible in 
experiments.

We note in passing that the effect of increasing the polymer concentration on 
the second osmotic virial coefficient of a globular protein-polymer solution
has been measured recently \cite{Kulkarni99}. However, these experiments are 
in the protein-limit $2 R_g / \sigma_{cc} \gg 1$, where the concepts of 
negative and positive non-additivity are less useful.

\acknowledgements

We thank M. Dijkstra, J. van Duijnenveldt, J.-P. Hansen, J. R. Henderson, 
H. L{\"o}wen, R. van Roij, G. Vliegenthart for illuminating discussions. RR 
acknowledges support from the EPSRC under grant No. GR/L89013, AAL 
acknowledges support from the Isaac Newton Trust, Cambridge.

\appendix
\section*{Low density limit of $\omega_1$}

Here we demonstrate that the one-body term $\beta \omega_1(z_s)$, 
given in Eq.~(\ref{cbbulk}), reduces in the limit $\rho_s^r \to 0$ to the 
{\em exact} low density limit, i.e.
\begin{equation} \label{wlimit}
\lim_{\rho_s^r\to 0}\beta \omega_1(z_s) = \rho_s^r V_{b+s},
\end{equation}
with $V_{b+s}=4 \pi \sigma_{bs}^3/3$ the volume of a spherical cavity of 
radius $\sigma_{bs}$ which is excluded to the centers of the small spheres. 
Recall that $\omega_1(z_s)$ is the excess chemical potential of a big hard
sphere in a sea of small ones.

In order to take the low density limit we note that to leading order in powers
of $\rho_s^r$ the equation of state, Eq.~(\ref{eos}), reduces to
\begin{equation}
\lim_{\rho_s^r\to 0} \beta p = \rho_s^r,
\end{equation}
the surface tension, Eq.~(\ref{gamma}), to
\begin{equation}
\lim_{\rho_s^r\to 0} \beta \gamma = \frac{\sigma_{ss}}{2} \rho_s^r,
\end{equation}
the coefficient of the term in  $\beta \omega_1$ linear in $R_b$ reduces to
\begin{equation}
\lim_{\rho_s^r\to 0} \frac{6 \eta_s^r}{\sigma_{ss} (1-\eta_s^r)} = \pi 
\sigma_{ss}^2 \rho_s^r
\end{equation}
and, finally,
\begin{equation}
\lim_{\rho_s^r\to 0} -\ln(1-\eta_s^r) = \frac{\pi}{6} \sigma_{ss}^3 \rho_s^r.
\end{equation}
It follows that
\begin{equation}
\lim_{\rho_s^r\to 0} \beta \omega_1 = \frac{4 \pi}{3} R_b^3 \rho_s^r + 
4 \pi R_b^2 \frac{\sigma_{ss}}{2} \rho_s^r + 4 \pi R_b 
\left(\frac{\sigma_{ss}}{2} \right)^2 \rho_s^r + 
\frac{4 \pi}{3} \left(\frac{\sigma_{ss}}{2}\right)^3 \rho_s^r.
\end{equation}
Recalling that $R_b\equiv \sigma_{bs}-\sigma_{ss}/2$ we find 
$\lim_{\rho_s^r\to 0} \beta \omega_1 = \frac{4 \pi}{3} \sigma_{bs}^3 \rho_s^r$
which is Eq.~(\ref{wlimit}).

\newpage

\end{document}